\documentclass[journal=jctcce,manuscript=article]{achemso}
\usepackage{xcolor}
\usepackage{color}
\author{Peng Bao}
\email{baopeng@iccas.ac.cn}
\author{Qiang Shi}
\email{qshi@iccas.ac.cn}
\affiliation[Chinese Academy of Sciences]
{Beijing National Laboratory for Molecular Sciences,
State Key Laboratory for Structural Chemistry of Unstable and
Stable Species, CAS Research/Education Center for Excellence
in Molecular Sciences, Institute of Chemistry, Chinese Academy
of Sciences, Zhongguancun, Beijing 100190, China}
\alsoaffiliation[University of Chinese Academy of Sciences]
{University of Chinese Academy of Sciences, Beijing 100049, China}
\author{Jiali Gao}
\email{gao@jialigao.org}
\affiliation[Beijing University Shenzhen Graduate School]
{Shenzhen Bay Laboratory, and Lab of Computational Chemistry and Drug Design,
Peking University Shenzhen Graduate School, Shenzhen, 518055, China}
\alsoaffiliation[University of Minnesota]
{Department of Chemistry and Supercomputing Institute, University of
Minnesota, Minneapolis, Minnesota 55455, United States}
\title[Block Localized Excitation]
{Diabatic States,Couplings and Potential Energy Surfaces through the
Block Localized Excitation Method}
\begin{document}
%%%%%%%%%%%%%%%%%%%%%%%%%%%%%%%%%%%%%%%%%%%%%%%%%%%%%%%%%%%%%%%%%%%%%
%% The manuscript does not need to include \maketitle, which is
%% executed automatically.  The document should begin with an
%% abstract, if appropriate.  If one is given and should not be, the
%% contents will be gobbled.
%%%%%%%%%%%%%%%%%%%%%%%%%%%%%%%%%%%%%%%%%%%%%%%%%%%%%%%%%%%%%%%%%%%%%
\begin{abstract}
%Diabatic electronic states are important to describe a variety
%problems involving quantum dynamics in molecular systems.
%In this article,
We propose a new block localized excitation (BLE) method
to directly construct diabatic excited states without the need to first compute
the adiabatic states.
%In the BLE method, the electronic localization is totally kept by orital
%localization, and there is no charge transfer but electrostatic, exchange
%and polarization between the blocks.
%The BLE method is based on the block-localized wavefunction approach,
The new method is capable to keep any electrons, spins,
and excitations localized in any divided blocks of
intermolecular and intramolecular systems. At the same time,
the electrostatic, exchange, and polarization interactions
between different blocks can be fully taken account of.
%To construct the excited states in the BLE method,
To achieve this, a new $\Delta$SCF
project method and the maximum wavefunction overlap method are
employed to obtain localized excited states with orbitals relaxation,
and the coupling between them are obtained using
approaches similar to the multistate DFT (MSDFT) method.
Numerical results show that the new BLE method is accurate
in calculating the electronic couplings of the singlet excitation
energy transfer (SEET) and triplet energy excitation transfer (TEET) processes,
as well as the excited-state intermolecular potential energy surface.
%The excited state and the ground state are orthogonal in the $\Delta$SCF
%project method.  For two face to face naphthalenes, the effective couplings
%of the singlet excitation energy transfer (SEET) and triplet energy excitation
% transfer (TEET) were studied by using direct couplings of BLE states.
%At the distances of 8-20 {\AA}, the coupling integral has good inverse
%relationship to the cube of the distance. The decay index of the couplings
%is xxx when Ms is equal to 1 and 0. A more accurate relation than the linear
%relation between the coupling and overlaps was also obtained.
%The coulomb coupling can be simply obtained from only two states
%when the overlap is small.
\end{abstract}
%%%%%%%%%%%%%%%%%%%%%%%%%%%%%%%%%%%%%%%%%%%%%%%%%%%%%%%%%%%%%%%%%%%%%
%% Start the main part of the manuscript here.
%%%%%%%%%%%%%%%%%%%%%%%%%%%%%%%%%%%%%%%%%%%%%%%%%%%%%%%%%%%%%%%%%%%%%
\section{Introduction}
Modern computational quantum chemistry usually starts with solving
%Schr\"{o}dinger equation. Adiabatic states are the direct solution of
the electronic Schr\"{o}dinger equation within the
Born--Oppenheimer (BO) approximation.\cite{born_zur_1927}
The resulted ground and excited states are adiabatic
states as they are the
eigenstates of the electronic Hamiltonian with fixed nuclear positions.
%including both the
%Adiabatic states are
%electronic wavefunction rapidly changes to adapt
%nuclear movement.\cite{born_zur_1927,born1998dynamical}
In cases where the BO approximation holds,
%the non-diagonal nuclear kinetic energy terms are small,
the adiabatic states provide good descriptions of the molecular
structure and interaction.
%can find the characteristics of the adiabatic states through experiments or
%computation directly. The non-diagonal nuclear kinetic energy terms in
%nuclear wave equation are corresponding to diagonal electronic Hamiltonians.
There are cases where the BO approximation becomes invalid,
typical examples include excited state dynamics near
avoided crossings or conical intersections,\cite{cederbaum2004conical,
koppel2004conical,baer2006beyond} as well as problems
of electron transfer (ET) and excitation energy transfer (EET)
reactions.\cite{may11} Many of these problems are not easily handled
using adiabatic states: Calculations of the non-adiabatic couplings
involve derivatives of the wave functions, and the derivative coupling
terms may often become singular at the crossing points.
% gives the
%singularity.\cite{cederbaum2004conical,koppel2004conical,baer2006beyond} In addition, in some research such as the couplings of
%electron transfer (ET) or excitation energy transfer (EET), the least energy
%differences between two adiabatic states need high accuracy computational
%methods in energy-gap-based method, and for large asymmetry systems it may
%be difficult to search for the resonance condition and the coordinate
%scanning may become computationally intensive.\cite{hsu2009electronic,you2014theory}
%\subsection{Diabatic state}

In contrast to the adiabatic states, diabatic states keep their physical
characters as the nuclear positions change, and the potential energy
surfaces are smooth.\cite{cederbaum2004conical,koppel2004conical,
baer2006beyond}
%The diagonal nuclear kinetic energy terms are corresponding to
%non-diagonal electronic Hamiltonians.\cite{cederbaum2004conical,koppel2004conical,baer2006beyond}
Moreover, the coupling between diabatic states does not involve
the derivate coupling terms due to the nuclear kinetic energy operator.
%the diabatic states are not eigenstates of the electronic Hamiltonian.
Usually, strict diabatic states cannot be obtained,\cite{mead1982conditions}
and many different methods are developed to construct
approximate diabatic states.\cite{hsu2009electronic,you2014theory,
van2010diabatic,subotnik2008constructing}
These methods can be classified into
two different categories. The first category of methods are based on
transformations from adiabatic states.
This class of methods %based on transformations from adiabatic states
can get good results, but the computational costs are relatively
high.\cite{hsu2009electronic,van2010diabatic,subotnik2008constructing}
In the second category of methods,
diabatic states are constructed directly by keeping their localized
character.\cite{van2010diabatic}
%The first class includes directly
%minimizing derivative couplings\cite{baer2006beyond,baer1975adiabatic,baer1980electronic}, designing diabatic states with
%intuitively desirable mathematical
%characteristics\cite{pacher1988approximately,
%pacher1993adiabatic,nakamura2001direct,nakamura2002direct,thiel1999proposal,
%koppel2001construction,ruedenberg1993quantum,atchity1997determination,
%pavlovic1987variational,domcke1993direct,domcke1994diabatic} and invoking a
%physical observable such as the dipole operator.\cite{mulliken1952molecular,
%hush1967intervalence,cave1996generalization,cave1997calculation,mcconnell1961intramolecular,
%onuchic1990predictive,kurnikov1996abinitio,beratan1998electron,stuchebrukhov1994dispersion,
%larsson1994calculation,hsu1997sequential,levy1992photophysics,voityuk2002fragment,
%pettersson2006singlet,hsu2008characterization,you2010fragment,werner1981mcscf,
%boys1960construction,subotnik2008constructing,subotnik2009initial,arago2015dynamics}
%Recent reviews on different methods to construct the diabatic states
%can be found in Refs.~\citenum{hsu2009electronic,you2014theory,
%van2010diabatic,subotnik2008constructing}.
%The directly constructing class also includes two kinds of methods.
There are different methods to directly construct diabatic states in the literature.
For example, the symmetry-broken method based on the
unrestricted Hartree-Fock (UHF)
\cite{bagus1972localized,broer1998hole} is found to
yield good results with a low computational cost. But they
are usually limited to singlet and triplet ground states and it may
be difficult to converge to the localized states.\cite{hsu2009electronic,you2014theory}

A different class of methods that attract many recent interests
construct the diabatic states based on the concept of fragmental
electron constraints. Among them, the constrained DFT (CDFT) method
is based on fragmental population constraints. The
diabatic states are constructed by CDFT through constraining the sum of
electron density to specific numbers in a certain
domain,\cite{dederichs1984ground,wu2005direct,kaduk2012constrained}
and a Lagrange multiplier is added to the whole Kohn-Sham
potential in CDFT.\cite{dederichs1984ground,wu2005direct,kaduk2012constrained}
CDFT can also be used to estimate ET and triplet
excitation energy transfer (TEET) couplings (Ref.~\citenum{kaduk2012constrained}
and references there in). Recently, CDFT was employed with non-Aufbau occupation
of the orbitals by the $\Delta $SCF method to construct localized singlet
excited state in singlet fission.\cite{yost2014transferable} One problem
of CDFT is that, the fragment domains need be selected carefully with
different population analysis methods since there is no exact partition.
Also, although the electron number in a certain domain can be constrained,
some electrons may not be localized in the specific domain, which may lead
to error diabatic states.\cite{mavros2015communication}
Subsystem DFT, or frozen DFT, is the another method based on fragmental orbital constraints.
The diabatic states are constructed through adding embedding
potential to Kohn-Sham potential of fragmental SCF.\cite{senatore1986density,
wesolowski1993frozen,jacob2014subsystem} Because of the approximation of the
non-additive part of the kinetic energy, Subsystem DFT is usually applied to weakly
interacting system.\cite{cembran2009block,jacob2014subsystem}

The valence bond (VB) theory %or non-orthogonal localized molecular orbitals
%({\color{red} does the NLMO equivalent to the VB theory ??? For one localized state,
%NOLMO can use one determinant but VB use lots of determinants})
naturally has the form of orbital constraints.
%The resonance
%structures in the VB
%theory have the concept of diabatic states as soon as VB theory
%appeared.\cite{pauling1960nature}
In the VB theory, the reactant and product states and
other resonance structures are represented by diabatic
states.\cite{pauling1960nature,shaik1981happens,shaik1999vb,
shaik2007chemist,song2008construction}
%It is difficult to
%get these VB diabatic states.\cite{song2008construction}
The {\it ab initio} VB theory is an appropriate
method to construct diabatic states,\cite{song2008construction,
cooper2009modern,goddard1973generalized,mcweeny1989valence,
van1991convergence,song2009efficient,
wu2011classical,chen2013nonorthogonal} but it can not handle the
large number of VB configurations for large and condensed phase
systems.\cite{song2008construction} To combine the advantages
of the molecular orbital (MO) theory and the VB theory,
a mixed molecular orbital and valence
bond (MOVB) theory\cite{song2008construction,mo2000abmovb,mo2000abcrs,
song2008effective,cembran2010non,gao2010generalized} and its
density functional theory (DFT) extension, the multistate
DFT (MSDFT) method,\cite{cembran2009block,mo2011energy} were developed, based on a
block-localized wavefuction (BLW) method\cite{mo1998theoretical,
mo1999simple,mo2000energy} and a similar
block-localized DFT (BLDFT) method,\cite{mo2007block} respectively.
There are other similar self-consistent field (SCF) localized methods
such as Stoll's method\cite{stoll1980use}, the self-consistent
field for molecular interactions (SCF-MI)\cite{gianinetti1996modification,
gianinetti1998extension,fornili2003determination}, locally projected MO (LPMO),\cite{nagata2001basis,
ferenczy2009optimization} localized MOs (ALMOs),\cite{khaliullin2006efficient,khaliullin2007unravelling}
and the gradient method with respect to orbital coefficients\cite{song2009efficient,
stoll1980use,smits1985calculation,couty1997extremely,sorakubo2003non}.

In the MOVB and MSDFT methods, the MOs and electrons are strictly localized
in the individual fragments in a diabatic state.
Therefore, the block-localized method can be considered as orbital
constraints method. There is no charge transfer but electrostatic,
exchange and polarization between different fragments.
The energy of the localized electronic state is variationally minimized through
SCF or direct minimization using gradient with respect to orbital
coefficients. Therefore, the block-localized method is computationally
inexpensive.
%, just like Hartree-Fork (HF) or DFT.
%The MOs are orthogonal using SCF and nonorthogonal using direct
%minimization in each fragment but
%nonorthogonal between fragments. The diabatic wavefuction is anti-symmetry
%
Recently, the block-localized diabatic state are applied
in many areas, mostly for ground state calculations in individual
fragments,\cite{cembran2009block,mo2011energy,mo2012block,
ren2016multistate} including two triplet ground state with
opposite spin in the recent study of singlet fission.\cite{chan2013quantum}
To obtain localized excited states, several methods were developed.
The simplified approach is to directly combine wavefunctions of
individual fragments.\cite{you2006triplet}
This approach ignores the interaction between
individual fragments including electrostatic, exchange and polarization. A
further simplified approach is the combination of the frontier orbitals of
individual fragments.\cite{scholes1994electronic,fujita1996abinitio}

In this work, we extended the block-localized methods to treat electronic states
that involve excited states in individual fragments.
Similar to previous works on ground states, the main focus is to
obtain localized excited states for a given fragment.
In principle, such calculations can be done using the
inexpensive TDDFT\cite{runge1984density} and the single excitation
configuration interaction (CIS) methods. But the TDDFT and CIS
are multi-configuration methods that makes the following MSDFT calculations
more complicated.
%More over, the TDDFT also has problem for charge transfer, energy transfer,
%Rydberg and double excitation states.\cite{tozer2000determination,
%dreuw2004failure}

We thus resort to single configuration methods and employ the
$\Delta $SCF method\cite{ziegler1977calculation} as the underlying method to
construct localized excited states.
%In the $\Delta $SCF method,
%the updated occupy orbitals are chosen from the total orbitals in every SCF
%iteration according to the same needed permutation of the ground state orbital
%energies order.
One problem of the original $\Delta $SCF method is that, the converged
excited states may collapse to the ground state or other excited states.
To solve this problem, the maximum overlap method (MOM),\cite{gilbert2008self}
and more recently, the maximum overlap square\cite{liu2014photoexcitation,
schmidt1993general} have been proposed.
%A local excitation $\Delta $SCF method was
%developed.\cite{ferre2002application}
%There is also the local SCF method for core-excited
%states.\cite{gavnholt2008delta}
%A self-consistent-field constricted variational DFT (SCF-CV-DFT) was
%suggested by constraining one electron to transfer from the occupied
%to the virtual orbital space.\cite{ziegler2009relation}
Once the single configuration localized excited states are obtained, further
construction of the diabatic states are rather similar to the
corresponding ground state MSDFT method.

The remainder of this paper is organized as follows. In
section 2, we present the theory and computational details of
the new BLE method. To obtain the
single-reference locally excited state, we also propose a new and efficient
$\Delta $SCF project method. Numerical results are then presented in
section 3, where we apply the new method to calculate SEET and TEET
couplings for a model system, as well as excited state intermolecular interactions
and excited state potential energy surface.
%couplings. At last, we study the electronic coupling of the excitation energy transfer (EET) using BLE method.
%presented in Sec. IV to demonstrate the effectiveness of the new
%method and to analyze the general properties of the high order expansion terms.
Finally, conclusions and discussions are made in section 4.

\section{Theory and Computational Method}
\subsection{Block-localized method for the ground state}
For completeness, we begin with a brief review of methods based on
molecular fragment or block localization, called block-localized wavefunction (BLW),
for the ground electronic state calculations. Similar approaches will be
used later to construct the diabatic states involving excited states.
In BLW, the determinant wave function is written as
follows:\cite{cembran2009block,mo1998theoretical,mo1999simple,mo2007block}
\begin{equation}
\Psi _u =N_u \mathop {\rm A}\limits^\wedge \left\{ {\Phi ^1\Phi ^2...\Phi
^K} \right\} \label{eqn:psiu}
\end{equation}
where ${N_u}$ is the normalization constant, $\mathop {\rm A}\limits^\wedge$
is the anti-symmetrization operator, $K$ is the number of molecular fragments,
or blocks, which could be a group of atoms, or a list of atomic orbital basis functions.
In Eq.~\ref{eqn:psiu}, $\Phi^A$ is a product of block-localized spin-orbitals,
$\Phi^A=\varphi_1^A \alpha \varphi_1^A \beta
...\varphi_{n_{A\alpha } }^A \alpha
\varphi_{n_{A\beta } }^A \beta $ in block A.
The orbitals from different blocks are generally nonorthogonal
since they typically do not share the same basis functions as a result of
block fragmentation.\cite{stoll1980use} The block-localized molecular orbitals (BLMO) are linear combinations
of atomic orbitals (LCAO) $\chi _{A\mu } $ within a specific block A:
\begin{equation}
\left| {\varphi_i^A } \right\rangle =\chi_A T_i^A =\sum\limits_{\mu =1}^{m_A
} {\left| {\chi_{A\mu} } \right\rangle } T_{\mu i}^A ,\qquad \qquad
\it A=1,2,...,K \label{eqn:phiiA}
\end{equation}
where block $A$ consists of $n_A$ $n_{A\alpha }+n_{A\beta }$ electrons
and $m_{A}$ basis functions, and the MO coefficients are denoted as $T_{\mu i}^A $.

The total coefficient matrix $T$ is block-diagonal when different blocks do not overlap,
but in the present approach, different blocks can share a group of common basis functions.
%where the coefficient matrix $T$ could be non-diagonal blocks or diagonal
%blocks when different blocks have the same basis sets in different blocks
%are overlap or not, respectively.
The total density matrix $D$ is given by
\begin{equation}
\label{eqn:eq1a}
D=T\left( {T^\dag ST} \right)^{-1}T^\dag
\end{equation}
where $S=\chi ^\dag \chi $ is the overlap matrix of the atomic
orbitals. The density matrix $D$ defined above satisfies the
symmetry ($D^{\dag } = D)$ and generalized idempotency (\textit{DSD = D})
conditions. The corresponding electron density is:
\begin{equation}
\label{eqn:eq1b}
\rho =\chi D\chi ^\dag \;\;.
\end{equation}

In the HF theory, the electronic energy is given by
\begin{equation}
\label{eqn:eq1c}
E= {\rm Tr}\left[ {D\left( {h+F} \right)} \right] \;\;,
\end{equation}
where $h$  is the one-electron core Hamiltonian,
and $F$ is the Fock matrix. In KS-DFT, the electronic energy is
\begin{equation}
\label{eqn:eq1d}
E= {\rm Tr}\left( {Dh} \right)+\frac{1}{2} {\rm Tr}\left( {DJD} \right)
+E_{xc} (\rho)
\end{equation}
where $J$ is the Coulomb integral matrix, and $E_{xc} (\rho )$ is the
exchange-correlation energy. The first-order variation of the energy with
respect to the coefficient matrix $T$ is given by:\cite{song2009efficient,
stoll1980use,gianinetti1996modification,smits1985calculation,
couty1997extremely,sorakubo2003non}
\begin{equation}
\label{eqn:eq1e}
\delta E = {\rm Tr}(\delta D\cdot F)=2 {\rm Tr}\left[ {\left( {I-SD_\alpha }
\right)F_\alpha T_\alpha \left( {T_\alpha ^\dag ST_\alpha }
\right)^{-1}\delta T_\alpha ^\dag +\left( {I-SD_\beta } \right)F_\beta
T_\beta \left( {T_\beta ^\dag ST_\beta } \right)^{-1}\delta T_\beta ^\dag }
\right]
\end{equation}

The energy gradients with respect to orbital coefficients can then be
obtained using Eq.~\ref{eqn:eq1e}, based on which the Broyden-Fletcher-Goldfarb-Shanno
(BFGS) updating scheme is used to optimize the BLMO.\cite{liu1989limited}
The gradient optimization method can be applied to
closed shell, restricted open shell, and unrestricted open shell cases.
Both analytic orbital gradient and numerical orbital gradient
have been implemented in our program.
It is usually easy to get
converged result when there is no common basis sets between different
blocks. When there is overlap of basis sets between different blocks,
it has been found previously that second order gradients methods
using approximate Hessian further improve the
convergence.\cite{couty1997extremely,sorakubo2003non}
%({\bf ??, move to the appendix})

Three different forms of SCF equations have been described.
%Through the variation of the energy, we can obtain three set SCF eqations of
%all groups. Because of the detail derivations elsewhere, we do not give the
%derivation but equations and references.
In the first method, the whole system is first
%To obtain group $A$, the first method is
separated into two parts, the part that belongs to
block $A$ ($\rho _{\notin A}$),
and the remainder of the system ($\rho _{\notin A}$).
The SCF equations for block $A$
are then given by:\cite{cembran2009block,mo2000energy,mo2007block,
gianinetti1996modification,
nagata2001basis,ferenczy2009optimization,nagata2004perturbation}
\begin{equation}
\label{eqn:blw-o}
(1-\rho_{\notin A} )f(1-\rho _{\notin A} )
\left| {\varphi_i^A } \right\rangle
=(1-\rho_{\notin A} )\left| {\varphi_i^A }
\right\rangle \varepsilon_i^A \;\;,
\end{equation}
which can also be written in matrix form
\begin{equation}
\label{eqn:blw-om}
{F_A'}{T_A} = {S_A'}{T_A}{E_A} \;\;.
\end{equation}
In Eq.~\ref{eqn:blw-om} ${F_A'} = {({I_{Na}} - {D_{ \notin A}}{S_{Na}})^\dag }F({I_{Na}} - {D_{ \notin A}}{S_{Na}}) $
and ${S_A'} = {S_{aN}}({I_{Na}} - {D_{ \notin A}}{S_{Na}}) = {S_{aa}}
- {S_{aN}}{D_{ \notin A}}{S_{Na}} $ where
$I$ is the unit matrix, the subscript $a$ denotes the basis functions
of block $A$, the subscript $N$ indicates full
dimension of basis functions of the whole system, $E$ is the matrix of orbital energies, $\varepsilon$,
the electron density operator $\rho _{\notin A}$
and the density matrix $D_{\notin A}$ describe the
parts that does not belong to block $A$, which are defined as:
\begin{equation}
\label{eqn:eq1f}
\rho _{\notin A} =\chi D_{\notin A} \chi ^\dag
\end{equation}
\begin{equation}
\label{eqn:eq1g}
D_{\notin A} =T_{\notin A} \left( {T_{\notin A} ^\dag ST_{\notin A} }
\right)^{-1}T_{\notin A} ^\dag
\end{equation}
In Appendix A, we also provide a more direct derivation
and properties of Eq.~\ref{eqn:blw-o}.

The second method to obtain the localized orbitals in
block $A$ is to solve the eigenvalue problem by using a
Hermitian operator $\rho _A^x$:\cite{khaliullin2006efficient,
gianinetti1998extension}
\begin{equation}
\label{eqn:eq1h}
(1-\rho +\rho _A^x )f(1-\rho +\rho _A^x )\left| {\varphi _i^A } \right\rangle
=(1-\rho +\rho _A^x )\left| {\varphi _i^A } \right\rangle \varepsilon _i^A
\end{equation}
where
\begin{equation}
\label{eqn:eq1i}
\rho _A^x =\chi T\left[ {\left( {T^\dag ST} \right)^{-1}} \right]_{.A}
\left[ {\left( {T^\dag ST} \right)^{-1}} \right]_{A.} T^\dag \chi ^\dag
\end{equation}
In Eq.~\ref{eqn:eq1h}, $\rho =\rho _A^x +\rho _{\notin A} $ (Appendix A), the superscript $A$ denotes the occupy orbitals
of block $A$.

In the third method, the eigenvalue problem is solved by using
a non-Hermitian operator $\rho _A^s $,\cite{stoll1980use,khaliullin2006efficient}
\begin{equation}
\label{eqn:eq1j}
(1-\rho + {\rho_A^s} ^\dag )f(1-\rho +\rho _A^s )\left| {\varphi _i^A }
\right\rangle =\left| {\varphi _i^A } \right\rangle \varepsilon _i^A
\end{equation}
\begin{equation}
\label{eqn:eq1k}
\rho _A^s =\chi T\left[ {\left( {T^\dag ST} \right)^{-1}} \right]_{.A}
T_A^\dag \chi ^\dag
\end{equation}

The three methods are derived
from the equation of zero energy gradient,
%({\color{red}\bf add references??})
and their computational costs are essentially the same when the number of
the blocks is not large. In this work, the first method is used
for the ground state calculations, and is extended to the
localized orbitals for excited states (below).
%We use the first method to do calculation in the
%next section.
%({\bf ??, why he algorithms are described in two different places - DIIS})

To accelerate the convergence of the SCF, we employ the
direct inversion in the iterative subspace (DIIS) method\cite{famulari2001application}
by using the energy gradients
(Appendix A) as the error vectors and updating the
projected Fock matrix and the effective overlap matrix for each
block. We have also implemented a DIIS by updating
the Fock matrix and coefficients of the whole system.
Efficiencies of the two methods are found to be similar.

\subsection{Block-localized excitation}
The structure of the BLE method is similar to that presented
in the previous subsection for the ground state, the main
difference is that we need to obtain the localized
excited state of a block, for example, block A, using a
set of localized orbitals.
Traditional excited state method are usually
of multi-configurational nature, even for the simplest
excited state methods such as CIS and TDDFT.
Using a multi-configurational wave function in
one block to represent local excitation states brings
more complexity to the block-localized algorithm presented in the previous subsection.
To avoid this problem, we employ the $\Delta $SCF like method,\cite{ziegler1977calculation}
which is a single configuration excitation method,
to construct the localized orbitals and the locally excited states,
such that the framework presented in section 2.1 can be preserved
in the BLE method.
%method. The original $\Delta $SCF method's convergence problem is from not
%choosing the correct orbitals.
%The localized excitation is that some electrons in the part of the system
%are excited to the quantum state of the higher energy. How to construct
%localized excited state? The simple way is to calculate each block then
%combine together. But it cannot be applied for intra-molecular excitation,
%and it has the problem of no interaction between blocks.
%%%%%%%%%%%%%%%%%%%%%%%%%%5
%\subsection{New single configuration excitation method}
%\begin{center}
%\textbf{$\Delta $SCF project method}
%\end{center}

The $\Delta $SCF method solves the localized orbitals in a way that
is very similar to the ground state Hartree-Fock-Roothaan equation.
%and is solved iteratively.
It is known that the original $\Delta $SCF method\cite{ziegler1977calculation} often
suffers from SCF convergence, and when it converges, an unwanted
excited state, or the ground state could be obtained.
One reason for the difficulty is due to the use of an identical order to
permute occupied and virtual orbitals in each SCF iteration.
However, the order of orbitals may have changed during the SCF from the initial guess.
% by simply using the orbital energies level.
% problem is from not
%choosing the correct orbitals.
The convergence can be greatly
improved by using the maximum overlap method (MOM),\cite{gilbert2008self}
and SCF convergence could be further improved by using maximum
overlap squared.\cite{liu2014photoexcitation}.

A number of local excitation $\Delta $SCF methods have been reported,
\cite{ferre2002application,gavnholt2008delta,ziegler2009relation,de1982converging,
glushkov2008frontiers,peng2013guided,weeks1968use,huzinaga1971theory,
surjan2000orthogonality,richings2007variationally,morokuma1972extended,
tassi2013hartree,tassi2013double,theophilou2014charge,evangelista2013orthogonality,
derricotte2015simulation}
including the local SCF method for
core-excited states\cite{gavnholt2008delta} and a
constricted variational DFT (SCF-CV-DFT) by one electron excitation to the
virtual orbital space.\cite{ziegler2009relation}
The single determinant approch can also be achieved by
orthogonality constraints on the excited state.
The ``big shift" method was formulated by setting a
large value (more than 10$^{10}$ a.u.) to the
diagonal element of the Fock matrix corresponding to ground state orbitals
.\cite{de1982converging}
Similarly, an asymptotic projection method was suggested
by adding a infinite projector to the frozen orbital in the
Hamiltonian operator.\cite{glushkov2008frontiers}
A yet another alternative is the guided SCF
approach by transformation of the Fock matrix from the atom orbitals (AO) basis to
the excited state orbitals basis.\cite{peng2013guided}
Similar to the BLW method, but enforcing orbital orthogonality
between core and valence, or between different fragments,
\cite{huzinaga1971theory},
is the projection configuration interaction method.\cite{surjan2000orthogonality}

In this work, we adopt the $\Delta $SCF approach
and employ two techniques to ensure SCF convergence.
%\hspace{-2em}\textbf{A new $\Delta $SCF project method}
In the first method, we first project the $\Delta $SCF equations
to the orbitals of the ground state space in each iteration.
As an example, we study the case where one $\alpha $ orbital is excited.
The ground state wavefunction is obtained from HF
or DFT, by solving the  Hartree-Fock-Roothaan equation iteratively:
\begin{equation}
\label{eqn:eq1l}
F^0T^0=ST^0E^0
\end{equation}
%where the orbital coefficients are given by $T_\alpha ^0 $.
The optimized ground state $\alpha $ orbitals are
$\left| {\varphi _{1\alpha }^0
\varphi _{2\alpha }^0 ...\varphi _{i\alpha }^0 ...\varphi _{n\alpha }^0
...\varphi _{i'\alpha }^0 ...\varphi _{m\alpha }^0 } \right\rangle $,
where $i$ denotes hole orbital (an occupied orbital), $i'$ denotes particle orbital
(an unoccupied orbital),
{$n\alpha $} is the number of {$\alpha $} electrons and $m$ is the basis
size. If we want to obtain the {$\alpha $}
excitation $\varphi_{i\alpha } \buildrel \over
\longrightarrow \varphi_{i'\alpha } $, the initial guess for orbitals of the
excited state is the $i\buildrel \over \longleftrightarrow i'$
permutation of the ground state {$\alpha $} orbitals,
$\left| {\varphi _{1\alpha }^0 \varphi _{2\alpha }^0 ...
\varphi _{i'\alpha }^0 ...\varphi _{n\alpha }^0 ...\varphi _{i\alpha
}^0 ...\varphi _{m\alpha }^0 } \right\rangle $, and there is no permutation of
the {$\beta $} orbitals.

By using the above initial guess, we start from the unrestricted open shell
Hartree-Fock-Roothaan SCF equation
\begin{equation}
\label{eqn:hfr}
f\left| {{\varphi }} \right\rangle  = \left| {{\varphi }} \right\rangle {\varepsilon }
\end{equation}
To solve this equation, we set $\left|\varphi\right\rangle =\chi T = \chi {T^0}T'$,
and left multiply $\left\langle {{\varphi ^0}} \right| = {\left( {\chi {T^0}}
\right)^\dag }$ %and in both sides of Eq.~\ref{eqn:hfr}
to obtain the matrix form of (Eq.~\ref{eqn:hfr}) in the ground state MO basis.
We have
\begin{equation}
\label{eqn:eq1m}
\left( {{T^0}^\dag F{T^0}} \right)T' = \left( {{T^0}^\dag S{T^0}} \right)T'E
\end{equation}
It is note that ${T^0}^\dag ST^0=I$ without localized constraint,
but it is not orthogonal in a localized block.
After solving the eigenstate problem by using Jacobi diagonalization, we obtain eigenvector $T'$ with
order $\left\{ {12...i...n...i'...m} \right\}$.
It is note that, after solving the eigenstate problem, the eigenvectors
are not sorted according to the energy of each orbitals as in
the conventional SCF calculation.
%without reorder according to orbital energie's values.
We then reorder $T'$ to $T'_{i\buildrel
\over \longleftrightarrow i'}$ with order $\left\{ {12...i'...n...i...m}
\right\}$ to construct the Fock matrix to realize $\Delta $SCF excitation,
%({\color{red}\bf ?? check....})
%Finally, when the SCF iteration converges, we obtain $T_{new}
%=T^0T'_{i\buildrel \over \longleftrightarrow i'} $ in AO basis to do next iteration.
i.e., $T_{new}=T^0T'_{i\buildrel \over \longleftrightarrow i'} $ in AO basis
is used to do the next iteration.
Alternatively, we can also use a permutated $T^{0}$ to realize the
$\Delta $SCF excitation, then follow above process without eigenvector
permutation. Usually, good convergence can be obtained by using projection of ground state orbitals.
We can also use the orbitals produced in previous iteration to do projection as a backup.
%({\color{red} \bf ??, what's the meaning of this sentence?? delete it???})

In a similar way, we can do any $\alpha $ and $\beta $ excitation
including multi-excitation through $\Delta $SCF project method. We note that
the guided SCF method\cite{peng2013guided} also employ a similar
projection method.
% using permuted occupied orbitals of the ground state
%after the eigenvectors are obtained. But it is more complicated than the
%$\Delta $SCF project method presented above, with the advantage that the
%excited state is orthogonal to the ground state Slater determinant.
The method presented above is not orthogonal, but the overlap between ground state and
excited state is usually very small.
To obtain orthogonal excited state,
the particle orbital can be obtained from the projection of unoccupied orbital space
after obtaining hole orbital from full space projection.
Similar to the derivation for the block-localized formalism,
the detail derivation within localized constraint is presented in the Appendix A.
In comparison, the hole orbital is obtained only from
the occupied orbital space in the EHP method.\cite{morokuma1972extended,tassi2013hartree}
% The obtained excited Slater
%determinant is orthogonal to ground state Slater determinant, and the
%overlap integral is less than 0.001 by calculation. Through our practical
%calculation, this method is robust to easily get convergence.
%\begin{center}
%\hspace{-2em}\textbf{Maximum wavefunction overlap method}
%\end{center}

Another method to improve the convergence of the original $\Delta $SCF method
is based on the idea of maximum wavefunction overlap with an initial guess of
an excited state Slater determinant.
%method, to choose
%the right orbital.
We also take one $\alpha $ orbital excitation as example.
Like the $\Delta $SCF project method presented above, we first obtain an
initial guess of the single reference excited state, by using the $i\buildrel \over
\longleftrightarrow i'$ permutation of the ground state \textit{$\alpha $}
occupied orbitals, $\Phi ^{0e}=\left| {\varphi _{1\alpha }^0
\varphi _{2\alpha }^0 ...\varphi _{i'\alpha }^0 ...\varphi _{n\alpha }^0 }
\right\rangle $, (the occupied \textit{$\beta $} orbitals are
kept unchanged). The overlap integral of the new wavefuction $\Phi $ and $\Phi
^{0e}$ is
\begin{equation}
\label{eqn:eq1n}
\left\langle {{{\Phi ^{0e}}}}
 \mathrel{\left | {\vphantom {{{\Phi ^{0e}}} \Phi }}
 \right. \kern-\nulldelimiterspace}
 {\Phi } \right\rangle  = \left| {{T^{0e}}^\dag ST} \right| = \left| {\begin{array}{*{20}{c}}
   {{O_{{1^{0e}}1}}} &  \ldots  & {{O_{{1^{0e}}n}}}  \\
    \vdots  &  \ddots  &  \vdots   \\
   {{O_{{n^{0e}}1}}} &  \cdots  & {{O_{{n^{0e}}n}}}  \\
\end{array}} \right|
\end{equation}
where $T^{0e}$ and $T$ are the initial and new occupied
orbital coefficients, respectively, $O_{ij}$ is the orbital overlap integral.
To obtain the maximum overlap between $\Phi^{0e}$ and $\Phi$,
we need to select $n$ occupied orbitals from $m$ total ones after diagonalization.
In principle, we can calculate all different overlap integrals, and choose the
combination with the maximum overlap. But the total number of overlap calculations
would be too big ($C_m^n$). A small active space of the hole and particle orbitals can be chosen
in practice. Otherwise, to simplify the problem, we consider the square
of the overlap integral,
%({\color{red}\bf ??, is this a determinant???}) yes
\begin{equation}
\label{eqn:ov2}
{\left\langle {{{\Phi ^{0e}}}}
 \mathrel{\left | {\vphantom {{{\Phi ^{0e}}} \Phi }}
 \right. \kern-\nulldelimiterspace}
 {\Phi } \right\rangle ^2} = \left| {{{\left( {{T^{0e}}^\dag ST} \right)}^\dag }
\left( {{T^{0e}}^\dag ST} \right)} \right| = \left| {\begin{array}{*{20}{c}}
   {\sum\limits_i {O_{{i^{0e}}1}^2} } &  \ldots  & {\sum\limits_i {{O_{{i^{0e}}1}}{O_{{i^{0e}}n}}} }  \\
    \vdots  &  \ddots  &  \vdots   \\
   {\sum\limits_i {{O_{{i^{0e}}n}}{O_{{i^{0e}}1}}} } &  \cdots  & {\sum\limits_i {O_{{i^{0e}}n}^2} }  \\
\end{array}} \right|
\end{equation}
The Frobenius norm of the above matrix is:
\begin{equation}
\label{eqn:eq1o}
\left\| {T^{0e}}^\dag ST \right\|_F =\sqrt {\sum\limits_i^n {\sum\limits_j^n
{O_{i^{0e}j}^2 } } }
\end{equation}
Therefore, to obtain the largest Frobenius norm, we select orbitals by
choosing the highest $n$ values of $\sum\limits_i^n {O_{i^{0e}j}^2 }, j=1,2,...,m $,
this is just the maximum overlap square method.\cite{gilbert2008self,liu2014photoexcitation}
Obviously, this method ignores the off-diagonal matrix elements in the right
of the (Eq.~\ref{eqn:ov2}).%, and lead to not converge in some difficult case.
To get better convergence, we can select orbitals by choosing the $n$ biggest
eigenvalues of $\left( {T^{0e}}^\dag ST \right)^\dag \left( {T^{0e}}^\dag
ST \right)$ where $T $ is $ m\times m$ dimension, just like the procedure to
obtain nature orbitals through the density matrix.
%({\color{red}\bf?? what's the difference here....}) diagnal elements v.s. eigenvalues
%The maximum wavefunction overlap method cannot keep the ground state and the
%excited state orthogonal because there is no orthogonal restriction. The
%overlap between the ground state and the excited state is small, usually
%less than 0.1.\cite{gilbert2008self}

%%%%%%%%%%%%%%%%%%%%%%%%
%In the block-localized excitation (BLE) method, equation 10 can be
%simplified to,
%\begin{equation}
%F_A^{'} T_A = S_A^{'} T_A E_A \label{eqn:hfrss}
%\end{equation}
%where $F'$ and $S'$ are effective Fock matrix and Overlap matrix. This is an
%eigenequation,

With the above $\Delta $SCF project method and the
maximum wavefunction overlap method, localized orbitals for excited states
for a given block $A$ can be obtained, and the formalism developed
in the previous subsection 2.1 can then be extended to treat the
excited state. In practical calculations, the block can be a molecular
fragment, molecule, or supermolecule. Since single configuration is used,
the $\Delta $SCF method does not have TDDFT's problem in some systems
such as charge transfer excitation, core excitation and double excitation.
Unrestricted open shell treatment is used in both $\Delta $SCF project
and the maximum wavefunction overlap method. They are found
to give the same energy after the convergence.
%The energy gradient and the geometry optimization
%can also be done by using the ground state program.
%In the $\Delta $SCF project method,
%the obtained excited Slater determinant is orthogonal to ground state Slater
%determinant, and the overlap is less than 0.001 by practical calculation.
We thus use the block-localized $\Delta $SCF project excitation method to do
all calculations in the next section.
%Unrestricted open shell treatment is used in both $\Delta $SCF project
%method and the maximum wavefunction overlap method. In practical
%calculation, both methods have the same energy results when the correct
%convergence was obtained, and the energy gradient and the geometry
%optimization can be done by using the ground state program. Because of
%single configuration, this method do not have TDDFT's problem in some field
%such as charge transfer excitation, core excitation and double excitation.
\subsection{Diabatic excited states and the electronic coupling}
The above obtained excited states with excitations
localized within a block are diabatic states.
As in previous application of the MSDFT
method for ground state calculations,
\cite{cembran2009block,mo2011energy,ren2016multistate} in order to
obtain the %(approximate)
adiabatic states, we need to compute
the coupling between these local excited states.
%In previous applications, the localized orbitals are all obtained
%from solving the ground state SCF equations.
%With the above new BLE method, we can now do the same thing for the excited
%states.\cite{cembran2009block,mo2011energy}
For a pair of diabatic states $\Psi_u$ and $\Psi_w$ calculated
using the HF method, the electronic coupling element can be
calculated as:\cite{broo1990electron,
farazdel1990electric,broer1988broken,
petsalakis1984nonorthonormal,mo2000abcrs}
\begin{equation}
\label{eqn:hks}
H_{uw} =\left\langle {\Psi _u \left| H \right|\Psi _w } \right\rangle
=M_{uw} \left[ {{\rm Tr}\left( {D_{uw}^\dag h} \right)+\frac{1}{2} {\rm Tr}\left(
{D_{uw}^\dag JD_{uw} } \right)-\frac{1}{4} {\rm Tr}\left( {D_{uw}^\dag KD_{uw} }
\right)} \right]=M_{uw}F_{uw}
\end{equation}
%({\color{red}\bf ??, this is for HF only???})
where $u$ and $w$ denote the two diabatic states, $F_{uw}$ is
a 'normalized' Hamiltonian matrix element,
%({\color{red}\bf ?? what is normalized??})
$M_{uw}$ is the overlap integral between the two diabatic states, which
is calculated from the determinant of the orbital overlap matrix,
\begin{equation}
\label{eqn:eq1p}
M_{uw} =\left| {T_u^\dag ST_w } \right|
\end{equation}
and the transition density matrix $D_{uw}$ is defined as:
\begin{equation}
\label{eqn:eq1q}
D_{uw} =T_w \left( {T_u^\dag ST_w } \right)^{-1}T_u^\dag
\end{equation}
%%%%%%%%%%%%%%%%%%%

In calculating the above coupling elements, L\"{o}wdin first obtained
the expression of Hamiltonian matrix element between two nonorthogonal
Slater determinants with the first- and the second-order cofactors.\cite{lowdin1955quantum}
Then the biorthogonal orbitals and singular value decomposition were applied
to simply the calculation.\cite{amos1961single,king1967corresponding}
The approach was further developed using density matrix and basis
functions without integral transformation.\cite{broo1990electron,farazdel1990electric,
broer1988broken,petsalakis1984nonorthonormal,mo2000abcrs}
%%%%%%%%%%%%%%%%%%%%%

The coupling term is calculated using the
correlation energies for the two diabatic states, which can also
be approximated by the energy difference between BLDFT and
that of Hartree-Fock theory using BLKS orbitals.\cite{cembran2009block,ren2016multistate}
\begin{equation}
\label{eqn:eq1t}
H_{uw} \approx H_{uw}^{KS} +\frac{1}{2}M_{uw}^{KS} \left[ {E_u \left( {\rho
_u } \right)-E_u^{KS} +E_w \left( {\rho _w } \right)-E_w^{KS} } \right]
\end{equation}
In this approach, we do not have the problems of the transition density
functional. We also derive a new approximate approach
which can derive the above result and another similar coupling term in Appendix B.
The numerical results from these approximate approaches are almost the same
in all the examples presented in this work. All next calculation uses Eq.~\ref{eqn:eq1t}.
There is also another transition density functional approach described in Supporting information.

After all the Hamiltonian matrix elements $H_{uw} $ and overlap
matrix elements $M_{uw}^{KS} $ are obtained, we can calculate the
adiabatic states %or needed diabatic (???) states
within the MSDFT framework,
by solving the secular equation,
\begin{equation}
\label{eqn:eq1u}
HC=MCE
\end{equation}
Obviously, the MSDFT method obtains the dynamic correlation first
and the static correlation at a later stage. It is thus different
from CASPT2, where the static
correlation are obtained first and the dynamic correlation are treated
later. The MSDFT method also has the advantages that, a much smaller
number of configurations are needed, and the computational cost is
at the DFT level.

Within the MSDFT framework,
%As stated above, an important application of the MSDFT method is to construct
%diabatic states and obtain the
electronic coupling between the diabatic states can also be calculated.
\cite{cembran2009block,ren2016multistate}
This is especially useful in calculating the coupling constants in the
electron transfer and excitation energy transfer processes.\cite{may11}
The effective coupling term $H_{uw}'$ between the $u$ and $w$ states
%coupling between two diabatic states
can be calculated conveniently
through L\"{o}wdin's orthogonalization,
%where the Hamiltonian matrix
%$H$ turns to $H'$ in orthogonal basis,
%\begin{equation}
%\label{eq35}
%H'C'=C'E
%\end{equation}
%One can also derive the effective diabatic coupling $H_{uw}'$
%between $u$ and $w$ states,
\begin{equation}
\label{eqn:eq1v}
{H_{uw}^{'}} = \frac{H_{uw} -M_{uw} \left( {H_{uu} +H_{ww} }
\right)/2}{1-M_{uw}^2 }
\end{equation}

All the methods presented in this section are implemented
in a modified version of GAMESS(US).\cite{schmidt1993general}

%%%%%%%%%%%%%%%%%%%%%%%%%%%%
\section{Results and discussion}
%We then apply the BLE method presented above to calculated excited state
%properties .... The excitation energy transfer coupling constants, and the
%interaction potential between an excited state molecule and ground state
%molecule (the ?? complex) are calculated.
As stated above, one of the important applications of the block-localized
excitation (BLE) method
presented above is to calculate the electronic coupling in the
the excitation energy transfer (EET) process, in which the electronic
excitation energy is transferred from one part of the molecular system to another
part.\cite{speiser1996photophysics} The EET process plays an important role
in many artificial and natural molecular systems such as organic light-emitting
diodes, photovoltaics, and photosynthetic system.\cite{holten2002probing,
baldo1998highly,gratzel2003dye,
mullen2006organic,fleming1997femtosecond,van2006energy}
%Both the EET and the electron transfer are similar.\cite{marcus1985electron}
In the  weak coupling limit, the EET rate constant can be calculated using the
Fermi's golden rule\cite{dirac1927quantum,you2014theory}:
\begin{equation}
\label{eqn:eq1w}
k_{EET} =\frac{2\pi }{\hbar }\left| V \right|^2\delta (E_i -E_f )
\end{equation}
%({\color{red}\bf ??, need this equation???}
where $V$ is the electronic coupling, $E_{i}$ and $E_{f}$ are the energy of the
initial and final states, respectively.
%initial/final state energy.
Therefore, the electronic coupling $V$ is a key parameter to determine
the EET rate constants.
\begin{scheme}
\includegraphics{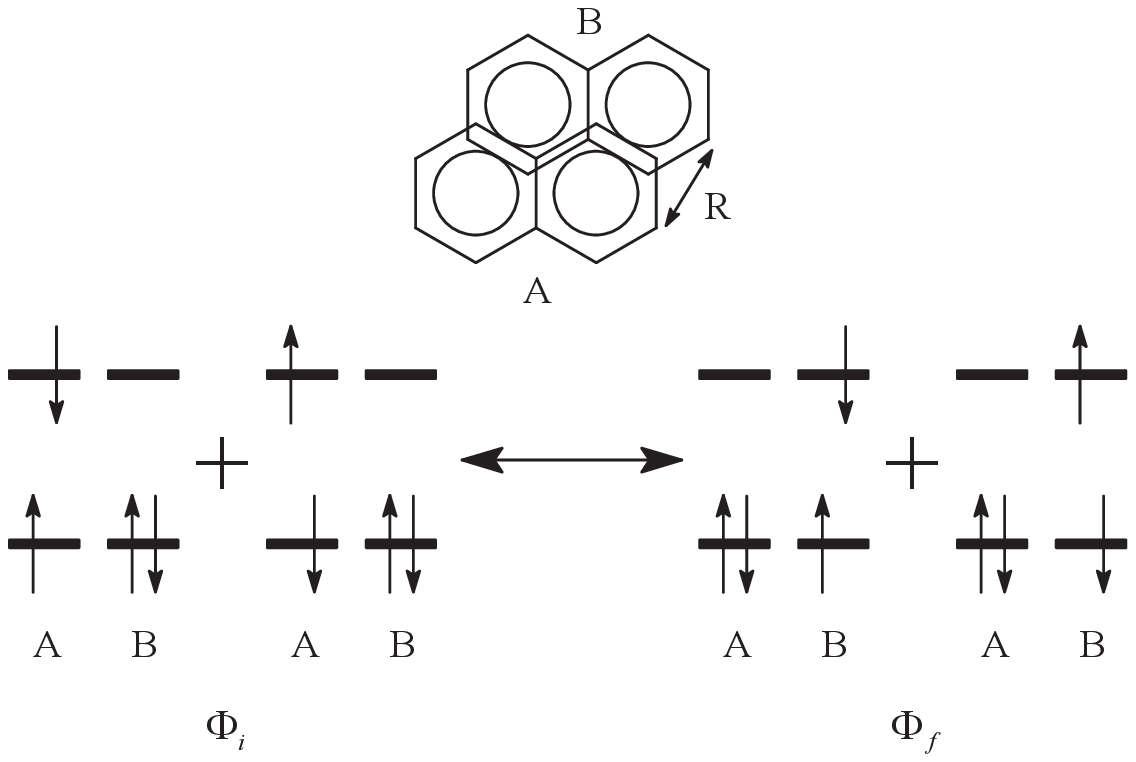}
\caption{The electronic configurations of HOMOs and LUMOs of
the initial and final states
in the SEET of the naphthalene dimmer.}
\label{sch:seet}
\end{scheme}
We apply the BLE method to calculate the EET couplings between two face to
face stacked naphthalene molecules (Scheme~\ref{sch:seet}),
which have been studied as a model system previously by using the combination of the frontier
orbitals of individual fragments\cite{scholes1994electronic} and
the fragment excitation difference (FED) method\cite{hsu2008characterization}.
The D$_{2h }$ geometry of the monomer was optimized at PBE0/6-31G* level in the
ground state, as we have assumed that the diabatic electronic coupling constant
does not change significantly with respect to the intramolecular degrees of freedom
(i.e., the Frank-Condon approximation is applied for the coupling constant).
All the calculations were carry out using unrestricted open shell at PBE0
level in a modified version of GAMESS(US).\cite{schmidt1993general}
In the BLE calculation, the dimmer is separated into two blocks, $A$
and $B$. Each block contains one naphthalene molecule,
whose HOMO and LUMO orbitals are displayed in Scheme~\ref{sch:seet}.
%{\{}Localized excitation energy, overlap integral calculation, test 3
%blocks, all not HOMO-LUMO orbitals are in one block{\}}

We first study the excitonic coupling in the
singlet excitation energy transfer (SEET) of the $1{}^1{B_{2u}}$
 excited state with HOMO to LUMO excitation.
The separation %with the change of the distance
between the two monomers ranges from 2.5 {\AA} to 20 {\AA}.
According to Scheme~\ref{sch:seet}, the localized
initial singlet excited state, $\Psi_i =\left| {A^\ast B} \right\rangle $,
is a spin-adapted excited state and linear combination of the two
localized excited states, $\Psi^{\alpha_A^\ast } $ and $\Psi^{\beta_A^\ast }$,
obtained from the BLE method with the HOMO to LUMO excitation
using the notation of the orbital wavefunctions.
%Because of the symmetry
%of this system, all orbitals of $\Psi^{\beta_A^\ast } $ are spin-flipped
%orbitals of $\Psi^{\alpha_A^\ast } $.
The final state $\Psi_f =\left| {AB^\ast } \right\rangle $ can be
obtained in a similar way, such that,
\begin{equation}
\label{eqn:eq1x}
\Psi_i =\frac{1}{\sqrt 2 }
\left( {\Psi ^{\alpha_A^\ast } +\Psi^{\beta_A^\ast } } \right)
\end{equation}
\begin{equation}
\label{eqn:eq1y}
\Psi_f =\frac{1}{\sqrt 2 }\left( {\Psi^{\alpha _B^\ast } +\Psi^{\beta_B^\ast } } \right)
\end{equation}

The lowest excited state ($1{}^1{B_{2u}}$) is a $\pi \to \pi ^\ast $ transition,
which has obvious single reference character.
% {\{}may test TD-B3LYP to know homo-lumo {\%} in adiabatic
%excitation state or just use ref.{\}},
We thus do not need more excited state configurations in calculating
the coupling constant.
%According to the symmetry characteristics,
Using symmetry simplification, Elements of the Hamiltonian and overlap matrices in
Eq.~\ref{eqn:eq1w} are then given by:
\begin{equation}
\label{eqn:eq1z}
H_{ii} =\left\langle {\Psi _i \left| H \right|\Psi _i } \right\rangle
=\left\langle {\Psi ^{\alpha _A^\ast } \left| H \right|\Psi ^{\alpha
_A^\ast } } \right\rangle +\left\langle {\Psi ^{\alpha _A^\ast } \left| H
\right|\Psi ^{\beta _A^\ast } } \right\rangle
\end{equation}
\begin{equation}
\label{eqn:eq2a}
H_{ff} =\left\langle {\Psi _f \left| H \right|\Psi _f } \right\rangle
=\left\langle {\Psi ^{\alpha _B^\ast } \left| H \right|\Psi ^{\alpha
_B^\ast } } \right\rangle +\left\langle {\Psi ^{\alpha _B^\ast } \left| H
\right|\Psi ^{\beta _B^\ast } } \right\rangle
\end{equation}
\begin{equation}
\label{eqn:eq2b}
H_{if} =H_{fi} =\left\langle {\Psi _i \left| H \right|\Psi _f }
\right\rangle =\left\langle {\Psi ^{\alpha _A^\ast } \left| H \right|\Psi
^{\alpha _B^\ast } } \right\rangle +\left\langle {\Psi ^{\alpha _A^\ast }
\left| H \right|\Psi ^{\beta _B^\ast } } \right\rangle
\end{equation}
\begin{equation}
\label{eqn:eq2c}
M_{if} =\left\langle {\Psi _i } \mathrel{\left| {\vphantom {{\Psi _i } {\Psi
_f }}} \right. \kern-\nulldelimiterspace} {\Psi _f } \right\rangle
=\left\langle {\Psi ^{\alpha _A^\ast } } \mathrel{\left| {\vphantom {{\Psi
^{\alpha _A^\ast } } {\Psi ^{\alpha _B^\ast } }}} \right.
\kern-\nulldelimiterspace} {\Psi ^{\alpha _B^\ast } } \right\rangle
+\left\langle {\Psi ^{\alpha _A^\ast } } \mathrel{\left| {\vphantom {{\Psi
^{\alpha _A^\ast } } {\Psi ^{\beta _B^\ast } }}} \right.
\kern-\nulldelimiterspace} {\Psi ^{\beta _B^\ast } } \right\rangle
\end{equation}
%By using equation 34 (because of some very small overlap integral, equation
%31 is not applied) and 37, the transition integral $V=\left| {H_{if}'}
%\right|$ can be obtained. The procedure is simplified to equation 27 and 37
%only with Kohn-Sham orbitals.

%({\color{red} \bf ??, Using symmetry simplification important, need to add some details of the BLE, otherwise it is not clear
%how the calculation are related to the above derivations...})
The effective excitonic coupling for the SEET can then be obtained using
Eq.~\ref{eqn:eq1t} and Eq.~\ref{eqn:eq1v}.
%({\color{red}\bf ??? below discusses general trends of the matrix elements,
%need to modify..??})
When the distance $R$ between the two monomers is larger than 10 {\AA},
the integral $\left\langle {\Psi^{\alpha_A^\ast } }
\mathrel{\left| {\vphantom {{\Psi ^{\alpha _A^\ast } } {\Psi
^{\alpha _B^\ast } }}} \right. \kern-\nulldelimiterspace} {\Psi ^{\alpha
_B^\ast } } \right\rangle $ is less than 10$^{-8 }$ because this is an
exchange integral just like TEET. When the distance $R$ between the dimers is
larger than 2.5 {\AA}, the integral $\left\langle {\Psi ^{\alpha _A^\ast } }
\mathrel{\left| {\vphantom {{\Psi ^{\alpha _A^\ast } } {\Psi ^{\beta
_B^\ast } }}} \right. \kern-\nulldelimiterspace} {\Psi ^{\beta _B^\ast } }
\right\rangle $ and $\left\langle {\Psi ^{\alpha _A^\ast } }
\mathrel{\left| {\vphantom {{\Psi ^{\alpha _A^\ast } } {\Psi ^{\beta
_A^\ast } }}} \right. \kern-\nulldelimiterspace} {\Psi ^{\beta _A^\ast } }
\right\rangle $ are less than 10$^{-8 }$ because one wavefunction
is just the spin-flip of the other in the same or different block.

\begin{figure}
\includegraphics[width=15cm]{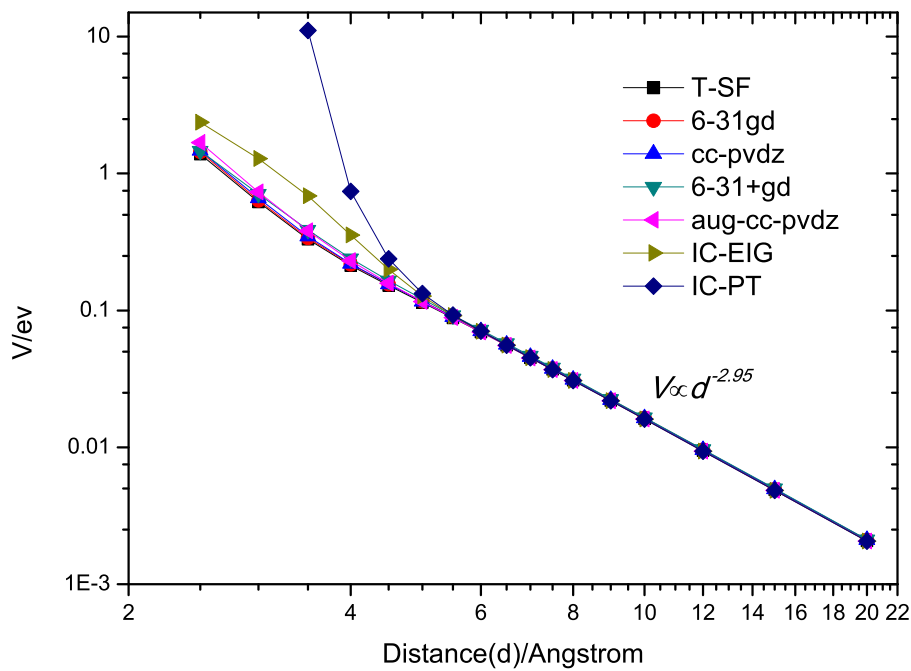}
\caption{The SEET coupling as a function of the intermolecular distance
for the $1{}^1{B_{2u}}$ state using PBE0 with different basis sets.
T-SF means that the singlet diabatic wavefunctions are from spin-flip
triplet block-localized wavefunctions ($M_s=1$) with 6-31G* basis set.
IC-PT means that the singlet diabatic wavefunctions include ionic configurations
$\left| {{A^ + }{B^ - }} \right\rangle $ and $\left| {{A^ - }{B^ + }} \right\rangle $
with the linear combination coefficients using the perturbation theory
at aug-cc-pvdz level.\cite{Harcourt1994rate}
IC-EIG means that the singlet diabatic wavefunctions include ionic configurations
$\left| {{A^ + }{B^ - }} \right\rangle $ and $\left| {{A^ - }{B^ + }} \right\rangle $
with the linear combination coefficients using eigen equation at aug-cc-pvdz level.
The fitted slope was for the distances between 8-20 {\AA}.}
  \label{fgr:seet}
\end{figure}
%{\color{red} \bf ?? important, use larger symbols, thicker lines and larger fonts...}}
%{\color{red}\bf ?? need more calculation, when CI after 6 {\AA}, why the lowest energy is
%only with two final states, need dzp+ like calculation, need optimize
%transition geometry by using constricted $\Delta $SCF project or BLE method;
%test V of 3,4,5,6... monomers}
%Through the configuration interaction equation 35, we can obtain the
%wavefunction of the adiabatic excited states, ${\left( {\Psi_i
%\pm \Psi_f} \right)} \mathord{\left/ {\vphantom {{\left( {\Psi_i \pm \Psi_f}
%\right)} {\sqrt {2\pm 2M_{if} } }}} \right. \kern-\nulldelimiterspace}
%{\sqrt {2\pm 2M_{if} } }$.

Figure~\ref{fgr:seet} shows the excitonic coupling as a function of
distance for the $1{}^1{B_{2u}}$ state. The results are similar to the ones obtained
using the FED method.\cite{hsu2008characterization} At distances larger than
about 6 {\AA}, the excitonic coupling is proportional to the inverse of
the cube of the distance, indicating that it is dominated by the
long range coulomb coupling, in agreement with the F\"{o}ster dipole-dipole
interaction.\cite{you2014theory,hsu2008characterization,forster1948zwischenmolekulare}
%has good inverse relationship to the cube of
%the distance. This is the
At distances smaller than about 6 {\AA}, the exchange coupling should also
contribute to the total excitonic coupling.
Different basis sets are used in the BLE calculation,
and it can be seen that there is only small differences between
the couplings calculated using different basis sets.

Further analyses of the SEET coupling are also performed in Figure~\ref{fgr:seet}.
The curve labeled T-SF shows the case where the singlet diabatic wavefunctions are obtained
from spin-flip of the triplet block-localized wavefunctions ($M_s=1$) with 6-31G* basis set.
The T-SF method gave the same couplings as the BLE method. The effect of ionic configurations
(ICs) was also considered because the ICs can be
the bridges from the initial to the final states.
IC-PT means that the singlet diabatic wavefunctions include ionic configurations (ICs)
$\left| {{A^ + }{B^ - }} \right\rangle $ and $\left| {{A^ - }{B^ + }} \right\rangle $
with the linear combination coefficients using the perturbation theory
at aug-cc-pvdz level.\cite{Harcourt1994rate,scholes1995rate,shi2012simplified}
IC-PT can only work when the coefficients of ICs are small. Therefore,
IC-PT should fail when the interaction between the two molecules is large
or the two molecules are too close. To get exact effects of ICs, the eigen equation of initial
and final states with ICs were also solved.
IC-EIG means that the singlet diabatic wavefunctions include ionic configurations
$\left| {{A^ + }{B^ - }} \right\rangle $ and $\left| {{A^ - }{B^ + }} \right\rangle $
with the linear combination coefficients using eigen equation at aug-cc-pvdz level.
Obviously, the excitonic couplings are overestimated by IC-PT method and
underestimated by the BLE method without ICs at distances smaller than about 4 {\AA}.
%In Figure~\ref{fgr:seet}, the difference of the basis sets have little effect on
%the SEET couplings.
%{\{}may add table of $\left\langle {\Psi ^{\alpha _A^\ast }
%\left| H \right|\Psi ^{\alpha _B^\ast } } \right\rangle ,\left\langle {\Psi
%^{\alpha _A^\ast } \left| H \right|\Psi ^{\beta _B^\ast } } \right\rangle
%$, equal in long range(no energy change for spin-flip) and $2\left\langle
%{\Psi ^{\alpha _A^\ast } \left| H \right|\Psi ^{\beta _B^\ast } }
%\right\rangle $ is coulomb coupling{\}}

In contrary to the SEET, excitonic coupling in triplet excitation energy
transfer (TEET) process of the $1{}^3{B_{2u}}$
excited state is dominated by short range coupling.
%We of T1 triplet state
%with the distances between two monomers from 2.5 {\AA} to 20 {\AA}.
According to Scheme~\ref{sch:seet}, like the SEET case studied above,
we can obtain two localized $M_s=0$ triplet excited states with HOMO to
LUMO excitation,
\begin{equation}
\label{eqn:eq2d}
\Psi_i =\frac{1}{\sqrt 2 }\left( {\Psi^{\alpha_A^\ast }
-\Psi^{\beta_A^\ast } } \right)
\end{equation}
\begin{equation}
\label{eqn:eq2e}
\Psi_f =\frac{1}{\sqrt 2 }\left({\Psi^{\alpha_B^\ast }
-\Psi^{\beta_B^\ast } } \right)
\end{equation}
The excitonic coupling of TEET with $M_s=1$ triplet states can also be
obtained by using diabatic states constructed by the
BLW method, which is shown in  Scheme~\ref{sch:teet}.
For simplicity, the electronic configurations for the $M_s=1$ states
are not shown.
\begin{scheme}
 \includegraphics[width=15cm]{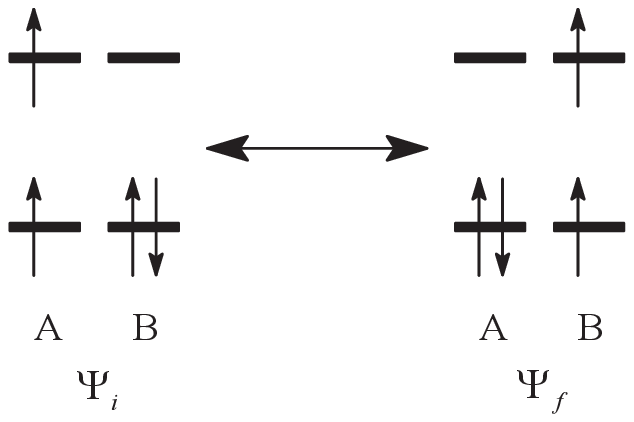}
  \caption{The electronic configurations of HOMOs and LUMOs of
the initial and final states
in the TEET of the naphthalene dimmer.}
  \label{sch:teet}
\end{scheme}

\begin{figure}
  \includegraphics[width=15cm]{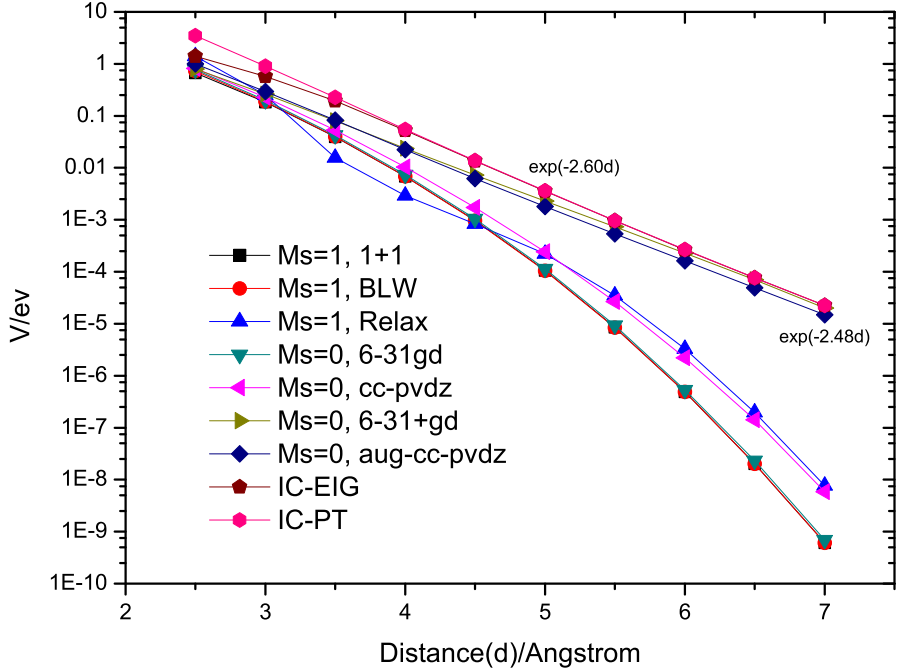}
  \caption{The coupling values vs. distances for the $1{}^3{B_{2u}}$ state using PBE0
with different basis sets."1+1" means the localized wavefunction is composed directly by
the wavefunctions of two blocks."Relax" means the wavefuctions are relaxed without any
restriction starting from BLW or "1+1".
"1+1","BLW" and "Relax" are all calculated with 6-31G* basis set.
IC-PT means the triplet diabatic wavefunctions of $M_s=0$ include ionic configurations
$\left| {{A^ + }{B^ - }} \right\rangle $ and $\left| {{A^ - }{B^ + }} \right\rangle $
with the linear combination coefficients using the perturbation theory
at aug-cc-pvdz level.\cite{Harcourt1994rate}
IC-EIG means the triplet diabatic wavefunctions of $M_s=0$ include ionic configurations
$\left| {{A^ + }{B^ - }} \right\rangle $ and $\left| {{A^ - }{B^ + }} \right\rangle $
with the linear combination coefficients using eigen equation at aug-cc-pvdz level.
the fitted exponents in the figures were obtained with ionic configuration IC-EIG
 (2.60{\AA}${^{-1}}$) at distances of 3.5-7 {\AA}
or without ionic configuration (2.48{\AA}${^{-1}}$) at distances of 2.5-7 {\AA}.}
  \label{fgr:teet}
\end{figure}
%Through the same progress in SEET, we obtain the transition integral of TEET
%with Ms=0 triplet state (Figure~\ref{fgr:teet}).

The calculated TEET excitonic couplings are shown in Figure~\ref{fgr:teet}
for both the $M_s=0$ and $M_s=1$ triplet states.
We first compare the "1+1","BLW", and "Relaxed" methods for $M_s=1$ states
and the BLE method for $M_s=0$ states with the 6-31G* basis set.
"1+1" means that the localized wavefunction is composed directly by
the wavefunctions of two blocks. "Relaxed" means that the wavefuctions
are relaxed without any restriction starting from BLW.
The couplings of "Relaxed" method are the same starting either from BLW or "1+1".
The couplings of "1+1", "BLW", and BLE methods have small
differences because the polarization interaction between two blocks is small
for the eclipsed naphthalenes at certain distances. The "Relaxed" method
did not give a smooth curve because the obtained wavefunction may deviate
from the most optimized state when the distance is less than 4.5 {\AA}.
Therefore, The "Relax" method may not be stable when the interaction between
two blocks are strong.

It is also shown in Figure 2 that,
the TEET couplings are not affected by different diffusive basis sets
6-31+G* and aug-cc-pvdz which decay exponentially with perfect correlation.
We can obtain the exponential decay of the TEET coupling
with the distance ${V_{TEET}} \propto {e^{ - \beta r}}$.
The exponential decay constant of the TEET couplings ${\beta}$ at aug-cc-pvdz level,
2.48 {\AA}${^{-1}}$, is similar to
the values found in Refs.~\citenum{hsu2008characterization,you2010fragment}.
However, couplings calculated using small basis sets without diffusive functions decay
faster than the ones with diffusive functions.

The ionic configurations are also found to have important effect on the TEET coupling
even at the distance 7 {\AA}, similar to the previous findings
in Ref.~\citenum{Harcourt1994rate,scholes1995rate,shi2012simplified}.
The couplings with ICs (IC-EIG and IC-PT) are about two times as the couplings without ICs.
The exponent of the IC-EIG TEET couplings at aug-cc-pvdz level, 2.60 {\AA}${^{-1}}$, is the same as
the exponent in Ref.~\citenum{hsu2008characterization}.
Obviously, compare with IC-EIG, the excitonic couplings are overestimated
by IC-PT method at distances smaller than about 3 {\AA}.
The exponential decay of the TEET coupling is similar to electron transfer coupling
because the TEET can be viewed as two electron exchanges with different spin.
\cite{hsu2009electronic} We further calculated the couplings of electron transfer (ET)
and hole transfer (HT) at aug-cc-pvdz level from 3.5 to 7 {\AA} using MSDDFT for
two napthalene molecules. The ${\beta}$ values of ET and HT are 1.00 and 1.45 {\AA}${^{-1}}$,
respectively. Thus, the ${\beta}$ value of TEET is almost the sum of the ${\beta}$ values
of ET and HT.
%{{V_{TEET}} = 0.479{ev^{ - 1}}{V_{ET}}{V_{HT}}} {R^{2}=0.9999}
%It is similar as Dexter exchange coupling.\cite{dexter1953theory}

According to Eqs.~\ref{eqn:hks} and \ref{eqn:eq1v},
when the overlap is small, the TEET coupling can be simplified to
\begin{equation}
\label{eqn:eq2f}
H_{uw}^{'} =\frac{M_{uw} }{1-M_{uw}^2 }\left( {F_{uw} -H_{uu} /2-H_{ww} /2}
\right)\approx M_{uw}C_{uw}
\end{equation}
Because the density or the orbital coefficients change slowly when varying the
molecular geometry in the diabatic states, $C_{uw}$ may be
approximate to a constant. The approximation $H_{uw}^{'} =CM_{uw} $ is first employed
in the extended H\"{u}ckel theory (EHT),\cite{hoffmann1963extended} and was applied in electron
transfer recently.\cite{troisi2002hole,cheung2010theoretical,gajdos2014ultrafast}
To evaluate the relationship between the TEET couplings and the overlaps for coupling values
spanning many orders of magnitudes(the least squares method cannot have the contribution of small value points),
the minimization of the statistical metric denoted exponential
root-mean-square logarithmic error (ERMSLE) was used.\cite{gajdos2014ultrafast,kubas2015electronic}
This means all the values of ${\frac{{H_{uw}'}}{{{M_{uw}}{C_{uw}}}}}$ should be approximately equal to 1.
\begin{equation}
\label{eqn:erms}
ERMLSE = \exp \left[ {\left\langle {{{\left( {\ln \frac{{H_{uw}'}}{{{M_{uw}}{C_{uw}}}}} \right)}^2}}
\right\rangle }^{1/2} \right]
%^{{\raise0.7ex\hbox{$1$} \!\mathord{\left/
% {\vphantom {1 2}}\right.\kern-\nulldelimiterspace}
%\!\lower0.7ex\hbox{$2$}}}}} \right]
\end{equation}
%({\color{red} \bf ??, add some descriptions of the above equation....})

\begin{figure}
  \includegraphics[width=15cm]{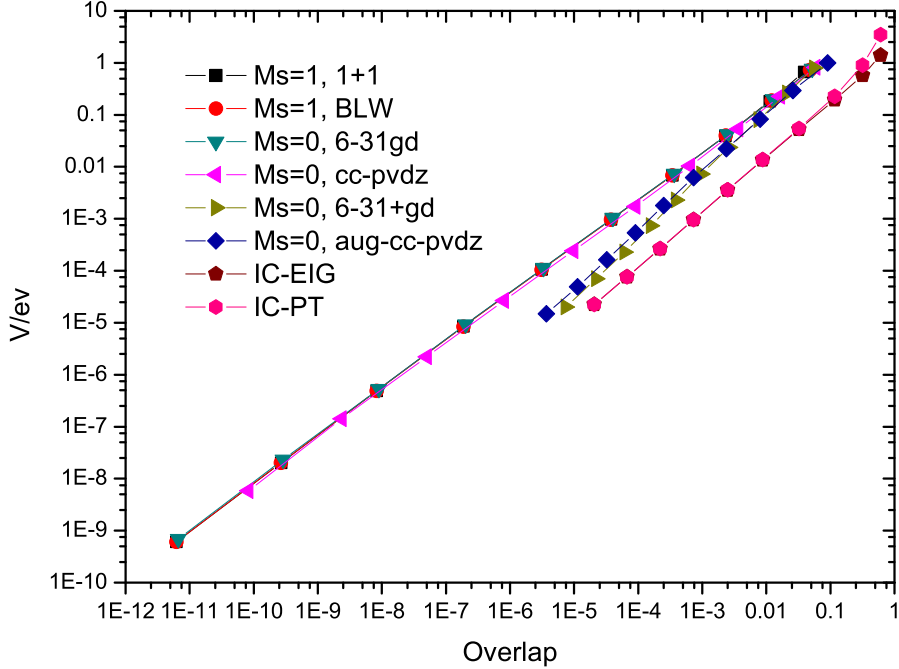}
  \caption{Correlations between the coupling constants and the wavefunction overlaps for TEET.
The labels for all the curves are the same as those in Figure~\ref{fgr:teet}.}
  \label{fgr:hcs}
\end{figure}

Figure~\ref{fgr:hcs} shows good linear correlation in the logarithmic scales
between the excitation couplings and overlaps for the dimmer with the
distances 2.5-7 {\AA}. We obtain the C values (7.22eV, $R^{2}$=0.992, ERMSLE=1.4
for $M_s=0$;
1.48eV, $R^{2}$=0.994, ERMSLE=1.2 for IC-EIG) at aug-cc-pvdz level.
From Figure~\ref{fgr:hcs}, we can also obtain a new relation $V=CM_{if}^{x }$
($V=17.0M_{if}^{1.11}$ eV, $R^{2}$=0.9994, $M_s=0$, aug-cc-pvdz;
$V=2.07M_{if}^{1.06}$ eV, $R^{2}$=0.9997, IC-EIG;
$V=10.5M_{if}^{0.908}$ eV, $R^{2}$=0.9999, $M_s=0$, 6-31G*).
%with very good correlations $R^{2}$=0.9999.
The reason of the relation is that both coupling and overlap of the TEET
decay exponentially with the distance.
%({\color{red}\bf ?? important, can this explain the above equation ???})
%(Figure in Support Information).
%Therefore, we can use diagonal matrix elements and overlap integral to
%approximate the non-diagonal matrix elements with Eq. (\ref{eqn:eq2g}) as an
%semi-empirical method, just like EHT but more accurate.
%Similarly, we can also approximate the correlation part of
%MSDFT non-diagonal matrix elements in future.
%\begin{equation}
%\label{eqn:eq2g}
%H_{uw} = M_{uw} \frac{\left( {H_{uu} +H_{ww} } \right)}{2}+CM_{uw}^x
%\end{equation}
%The coulomb coupling of the SEET can be obtained from the difference of the
%total coupling and the exchange coupling. We treat the effective coupling of
%Ms=1 TEET as the exchange coupling,
%\begin{equation}
%\label{eq48}
%{H_{uw}^C}'={H_{uw}^{SS}}'-{H_{uw}^{TT}}'=2\frac{\left\langle
%{\Psi^{\alpha_A^\ast } \left| H \right|\Psi^{\beta_B^\ast } }
%\right\rangle -M_{uw}
%\left\langle {\Psi^{\alpha_A^\ast } \left| H \right|\Psi^{\beta_A^\ast
%} } \right\rangle }{1-M_{uw}^2 }
%\end{equation}
%The coulomb coupling displays the obvious dipole-dipole $d^{-3}$ relation.
%Since the overlap integral $M_{uw} $ is very small when the distance is less
%than 3 {\AA}, the coulomb coupling is simplified to $2\left\langle {\Psi
%^{\alpha _A^\ast } \left| H \right|\Psi ^{\beta _B^\ast } } \right\rangle$.
%({\color{red}\bf ??, add results for the excimer interactions...})

The above calculations of SEET and TEET couplings concern Frenkel excitons,
where the electron and hole are localized within the same block.
The charge transfer states also play important roles in determining excited state
properties. In the following, we apply the new BLE method to calculate the
intermolecular interaction potential energy curve
in the napthalene excited dimer (excimer),
where the exciton resonance state (ER, ${A^ * }B \leftrightarrow A{B^ * }$)
and the charge resonance state (CR, ${A^ + }{B^ - } \leftrightarrow {A^ - }{B^ + }$)
are the origin of the strongly
attractive interaction.\cite{birks1975excimers}

Aromatic excimers, such as that formed between two naphthalenes, are often studied
because of their typical photophysical and photochemical characters.
With the development of computational chemistry, many different theoretial methods
have been applied to study aromatic excimers, including
the semiemperical,\cite{ azumi1964energy, chandra1968semiempirical}
TDDFT,\cite{kolaski2013aromatic} and post-SCF\cite{scholes1994electronic,
shirai2011ab, shirai2016computational} methods.
In these works, the calculated properties of the excimer, such as
binding energy, absorption energy, and emission energy, were found
to heavily depend on the methods and basis sets.
Up to date, the most advanced method to calculate the naphthalene excimer
is the DMRG-CASPT2 (the density matrix renormalization
group-the complete active space second order perturbation theory)
approach using the full $\pi$ valence orbitals as the active space,
because of strong static and dynamic correlation.\cite{ shirai2016computational}
In comparison, the block-localized excited MSDFT (BLE-MSDFT) is much
less computationally demanding, and is much easier to be applied to
larger systems.
%Obviously, we want to use less diabatic states to construct more accurate PES
%of the excimer through

\begin{figure}
\includegraphics[width=15cm]{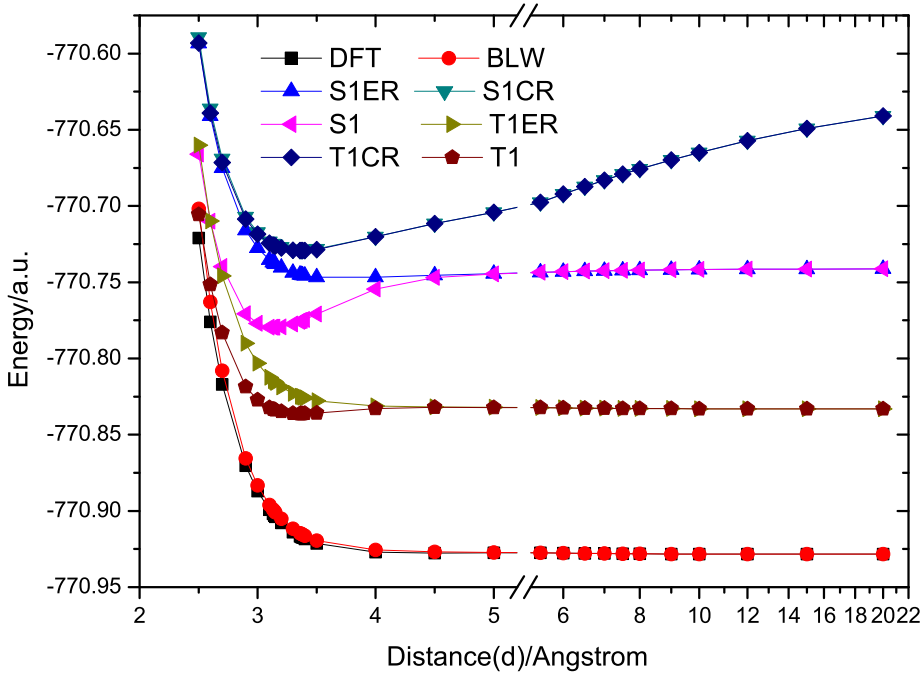}
\caption{Potential energy curves of the ground and excited states of the
naphthalene dimer using PBE0/cc-pvdz. DFT: the ground state energy using DFT; BLW:
the ground state energy using BLW
%({\color{red}\bf ??, with how many diabatic states???}: BLW has only one state like DFT)
; S1ER/T1ER: excitonic-resonance states,
the lowest singlet/triplet state energy of the CI of four localized singlet excited states
($\Psi^{\alpha_A^\ast } $, $\Psi^{\beta_A^\ast }$, $\Psi^{\alpha_B^\ast } $ and $\Psi^{\beta_B^\ast }$);
S1CR/T1CR: charge-resonance states, the lowest singlet/triplet state energy of the CI of four singlet
charge-transfer states; S1/T1: the first singlet/triplet excited state, the lowest singlet/triplet
state energy of the CI of four localized singlet excited states and four singlet charge-transfer states.}
\label{fgr:pes}
\end{figure}

To calculated the intermolecular interaction in the excimer state, we use
all the localized and charge-transfer states through HOMO-LUMO singlet excitation
to construct ER and CR states.
The singlet/triplet excited states are then determined as
combinations of ER and CR states using the MSDFT method.
The potential energy surface (PES) is shown in Figure~\ref{fgr:pes}.
%({\color{red} \bf ??, important, explain what are the different curves
%in Figure 4...})
DFT means the ground state energy using DFT. BLW means
the localized ground state energy using BLW. S1ER/T1ER means the excitonic-resonance states,
the lowest singlet/triplet state energy of the CI of four localized singlet excited states
($\Psi^{\alpha_A^\ast } $, $\Psi^{\beta_A^\ast }$, $\Psi^{\alpha_B^\ast } $ and $\Psi^{\beta_B^\ast }$).
S1CR/T1CR means charge-resonance states, the lowest singlet/triplet state energy of the CI of four singlet
charge-transfer states. S1/T1 means the first singlet/triplet excited state, the lowest singlet/triplet
state energy of the CI of four singlet localized excited states and four singlet charge-transfer states.
%The lower excited
%states of the dimer are also called excimer, and the Frenkel exciton is
%localized in a monomer.
%The excimer can be applied to many fields, such as excimer lasers, singlet fusion,
%molecular sensing and biophysics
%.\cite{ basov1970laser, chan2013quantum, xu2009unique, somerharju2002pyrene}
%There are many experimental,\cite{ birks1975excimers, saigusa1996excimer}
%theoretical and computational
%researches about excimers.\cite{azumi1964energy, chandra1968semiempirical,
%scholes1994electronic, kolaski2013aromatic, shirai2011ab, shirai2016computational}

We first analyze the singlet states. The excimer state of the naphthalene
is the lower state excited from HOMO to LUMO, named the $L_a$ state.\cite{shirai2011ab}
In contrast, for the monomer of naphthalene, the $L_b$ state, excited from HOMO to LUMO+1
and HOMO-1 to LUMO, is the lowest excited state, and the $L_a$ state
is the second lowest excited state. \cite{shirai2011ab}
In present work, only the $L_a$ excited state was considered when calculating
the excimer properties.

With the increase of the intermolecular distance, the S1 and S1ER energies
first decrease to a minimum and then increases.
The S1ER state has a very small binding energy compare to S1 state,
such that the S1CR states play an important role for the binding energy of the
eximer state. As shown in Figure 4, the energy of the S1CR state clearly show
the charactoristics of the Coulombic interaction between two opposite ions
at a distance larger than 4 {\AA}.\cite{azumi1964energy}

\begin{table}
\caption{Intermolecular Equilibrium Distance ($r_e$), Binding Energies (BE),
and Transition Energies of the Naphthalene singlet state excimer.
The numbers in parentheses are from BLW of ground state for eliminating BSSE.}
  \label{tbl:s1}
  \begin{tabular}{llllll}
    \hline
                &             &             &         & \multicolumn{2}{l}{Transition energy (eV)}  \\
    Method      & Basis set   & $r_e$ ({\AA}) & BE (eV) & $r=r_e$  & $r$=20 ({\AA}) \\
    \hline
    BLE-MSPBE0	& 6-31G(d)    & 3.11        & 0.99    & 3.40(3.31) & 5.16 \\
    BLE-MSPBE0	& 6-31+G(d)   & 3.19        & 1.07    & 3.36(3.28) & 5.07 \\
    BLE-MSPBE0	& cc-pVDZ     & 3.14        & 1.04    & 3.36(3.27) & 5.09 \\
    BLE-MSPBE0	& aug-cc-pVDZ & 3.18        & 1.10    & 3.29(3.19) & 4.99 \\
    BLE-MSPBE	& cc-pVDZ     & 3.14        & 0.95    & 3.11(3.01) & 4.75 \\
    BLE-MSB3LYP	& cc-pVDZ     & 3.20        & 0.79    & 3.46(3.38) & 5.03 \\
    BLE-MOVB	& cc-pVDZ     & 3.35        & 0.91    & 4.23(4.17) & 5.85 \\
    exptl.      &	      & 3.0-3.6\textsuperscript{\emph{a}} &
    0.76\textsuperscript{\emph{b}},\textsuperscript{\emph{c}} &
    3.13\textsuperscript{\emph{b}} & 4.45\textsuperscript{\emph{c}},
    4.7\textsuperscript{\emph{d}} \\
    \hline
  \end{tabular}

  \textsuperscript{\emph{a}} Ref.~\citenum{forster1969excimers};
  \textsuperscript{\emph{b}} Ref.~\citenum{azumi1964energy};
  \textsuperscript{\emph{c}} Ref.~\citenum{george1968intensity};
  \textsuperscript{\emph{d}} Ref.~\citenum{mcconkey1992excitation}.
\end{table}

Table~\ref{tbl:s1} presents the theoretical and experimental spectroscopic parameters
of the naphthalene excimer state using different kinds of methods and basis sets.
For the BLE-MSPBE0 method, the spectroscopic parameters
show little basis set-dependent by using different basis sets
6-31G(d), 6-31+G(d), cc-pVDz and aug-cc-pVDZ.
The intermolecular equilibrium distance, $r_e$, ranges from 3.11 to 3.19 {\AA},
which is consistent with the experimental value and previous theoretical
works.\cite{forster1969excimers, shirai2011ab, shirai2016computational}
The binding energies, BE, ranges from 0.99 to 1.10 eV,
is higher than the experimental value, and is consistent with
the ground state BSSE (basis set superposition errors) corrected
DMRG-CASPT2 values in Ref.~\citenum{shirai2016computational}.
The transition energies at the excimer structures, ranges from 3.29 to 3.40 eV,
is very close to the experimental value.
The transition energies at the superdimer structures, ranges from 4.99 to 5.16 eV,
is slightly higher than the experimental value.\cite{george1968intensity, mcconkey1992excitation}
% ({\color{red}\bf ??, add references....})
	
There is no BSSE by using BLW method because of separated basis sets.\cite{gianinetti1996modification}
Therefore, no BSSE is
in the binding energies by using MSDFT calculation with same partition as monomers.
The result in parentheses of the Table~\ref{tbl:s1} is in line with this reason.
At the equilibrium distance, the block-localized form eliminates the BSSE
 (0.11 eV/cc-pVDZ, 0.10 eV/aug-cc-pVDZ for the ground state) and
is consistent with the BSSE (0.11 eV/cc-pVDZ, 0.10 eV/aug-cc-pVDZ for the ground state)
by using the counterpoise procedure.
%({\color{red} \bf ??, the discussion of BSSE is not clear enough.....})
In contrast, the basis set-dependence is quite obvious
by using different basis sets from 6-31G(d) to aug-cc-pVDZ in Ref.~\citenum{shirai2011ab} .
Similarly, all the spectroscopic parameters show little dependece on the
density functionals in the BLE-MSDFT method (PBE, PBE0 and B3LYP) at cc-pVDZ level.
In contrast, the density functional-dependence is obvious
using different density functionals (PBE, PBE0 and B3LYP)
in Ref.~\citenum{kolaski2013aromatic}.
The severe deviation between the values of BLE-MOVB and the experimental values
indicates that the dynamic correlation is necessary
for the intermolecular interactions in the excited states.

Then we analyze the triplet excited state of the naphthalene.
There are still debates on the existence of the triplet naphthalene
excimer from both experimental and theoretical
studies.\cite{ langelaar1968monomer, takemura1976kinetic,birks1975excimers,
lim1987molecular, chandra1968semiempirical, pabst2011triplet, kim2015effects}
The PES of T1CR state is almost the same as the PES of S1CR state.
The T1ER state energy decreases monotonously as the distance increases.
Thus, the T1CR states play an important role
when the distance between two parts is less than 4 {\AA}.
The T1 state potential energy curve has a mimumum at around  3.4 {\AA},
and a maximum at about 5 {\AA}.
%and then continues monotonous declined very slowly as the distance increases.
There is 0.02-0.03 eV energy difference between the maximum energy
at about 5 {\AA} and the energy at 20 {\AA}.
The T1 state binding energy is about 0.1eV, which is much smaller than
the S1 state binding energy.

\begin{table}
\caption{Intermolecular Equilibrium Distance ($r_e$), Binding Energies (BE),
and Transition Energies of the Naphthalene triplet state excimer by
using BLE-MSPBE0. The numbers in parentheses are from BLW of ground state for eliminating BSSE.}
  \label{tbl:t1}
  \begin{tabular}{lllll}
    \hline
                &            &         & \multicolumn{2}{l}{Transition energy (eV)}    \\
    Basis set	& $r_e$ ({\AA}) & BE (eV) & $r=r_e$  & $r$=20 ({\AA}) \\
    \hline
    6-31G(d)	& 3.39	     & 0.02    & 2.29(2.21) & 2.60         \\
    6-31+G(d)	& 3.45	     & 0.07    & 2.26(2.20) & 2.60         \\
    cc-pVDZ	    & 3.37	     & 0.08    & 2.21(2.15) & 2.59         \\
    aug-cc-pVDZ	& 3.40	     & 0.12    & 2.21(2.13) & 2.59         \\
    exptl.      &            &         & 2.3\textsuperscript{\emph{a}} & 2.64\textsuperscript{\emph{b}} \\
    \hline
  \end{tabular}

  \textsuperscript{\emph{a}} Ref.~\citenum{langelaar1968monomer, takemura1976kinetic};
  \textsuperscript{\emph{b}} Ref.~\citenum{lewis1944phosphorescence}.
\end{table}

Table~\ref{tbl:t1} gave the experimental and theoretical spectroscopic parameters
of the naphthalene dimmers by using BLE-MSPBE0 method.
The spectroscopic parameters are very less basis set-dependent
by using different basis sets 6-31G(d), 6-31+G(d), cc-pVDz and aug-cc-pVDZ:
the intermolecular equilibrium distance, $r_e$, ranges from 3.37 to 3.45 {\AA};
the binding energies, BE, ranging from 0.02 to 0.12 eV,
are very small to reflect the argument of forming the triplet excimer;
the transition energies, 2.21-2.29 eV at equilibrium distance and 2.59-2.60 eV at 20 {\AA},
are very close to the experimental value, 2.3 eV and 2.64 eV, respectively.	
The block-localized form eliminates the BSSE (0.08 eV/aug-cc-pVDZ for the ground state) and
is consistent with the BSSE (0.08 eV/aug-cc-pVDZ for the ground state)
by using the counterpoise procedure.

\section{Conclusions}
In summary, we have proposed an efficient BLE method,
which combines the block-localized approach and the single
configuration excitation method to directly construct diabatic
excited states. Two new single configuration excitation methods,
the $\Delta $SCF project method and the maximum wavefunction overlap
method are developed and implemented in the BLE method.
%because they only treat
%the general eigenequation.
Within in the framework of MSDFT, we show how the new BLE method can
be applied to calculate the excitonic couplings in the SEET
and TEET processes.

Numerical results show that the new BLE method is accurate
in calculating the SEET and TEET coupling constants,
and the excited state intermolecular interactions in aromatic excimers.
The calculated results also show little dependence on the choice
of basis sets and density functionals.
Although the calculations are performed on model systems
consists of two identical monomers, the method can certainly
be applied to heterodimers, molecular fragments, and possibly more
complex molecular blocks. We thus expect that the new method
can be applied to many problems of excited state structure
and dynamics.
%More applications of diabatic excited states, construction of adiabatic states
%and dynamics on excited states in many fields can be done by keeping
%different electrons, spins and excitations localized in the future.

In the future, the BLE may be further simplified to use the
Hartree product wavefuction between different blocks, such as the
polarization of Morokuma energy decomposition analysis\cite{kitaura1976new},
explicit polarized potential (X-pol)\cite{gao1997toward}, fragment
molecular orbital(FMO),\cite{kitaura1999fragment} and restrict geometry
optimization for aromaticity (RGO)\cite{bao2011new} methods.
The Hartree product wavefunction as diabatic state is accurate when
the distance between two parts is large because the exchange and
repulsion between two parts are small and the polarization effect
is taken account of. It is also a linear scaling method for the
large system, and can be applied to systems such as large molecular
aggregates.

\begin{acknowledgement}
This work is supported by NSFC (Grant No. 21673246),
and the Strategic Priority Research Program of the Chinese Academy
of Sciences (Grant No. XDB12020300).
\end{acknowledgement}
%\begin{acknowledgement}
%\textbf{Appendix: the direct derivation of the energy gradient and SCF (Eq.~\ref{eqn:blw-om})}
%\begin{appendices}
\appendix
\renewcommand\thesection{\appendixname~\Alph{section}}
%\renewcommand\theequation{\arabic{equation}}
%%%%%%%%%%%%%%%%%%%%%%55
\section{Calculation of the block-localized ground and excited state energy gradients }
%%%%%%%%%%%%%%%%%%%%%%%%%%%%
The operator form of the ground state energy gradient of the equation 8 for
RHF is given by:\cite{stoll1980use}
\begin{equation}
\label{eqn:eq2h}
\frac{\partial E}{\partial \left\langle {{\varphi _A}} \right|}=4(1-\rho
)f\left| {\tilde {\varphi }_A } \right\rangle
\end{equation}
where the reciprocal orbitals $\left| {\tilde {\varphi } } \right\rangle$ are
\begin{equation}
\label{eqn:eq2i}
\left| {\tilde {\varphi }} \right\rangle =\left| {\tilde {\varphi }_A }
{\tilde {\varphi }_{\notin A} } \right\rangle
=\left| {\varphi _A \varphi _{\notin A} } \right\rangle
\left\langle {\varphi _A \varphi _{\notin A} }
\mathrel{\left| {\vphantom {{\varphi _A \varphi _{\notin A} }
{\varphi _A \varphi _{\notin A} }}} \right. \kern-\nulldelimiterspace}
{\varphi _A \varphi _{\notin A}
} \right\rangle ^{-1}=\left| {\varphi _A \varphi _{\notin A} } \right\rangle
\left( {{\begin{array}{*{20}c}
{\left\langle {\varphi _A } \mathrel{\left| {\vphantom {{\varphi _A } {\varphi _A
}}} \right. \kern-\nulldelimiterspace} {\varphi _A } \right\rangle } \hfill &
{\left\langle {\varphi _A } \mathrel{\left| {\vphantom {{\varphi _A }
{\varphi _{\notin A} }}} \right. \kern-\nulldelimiterspace} {\varphi _{\notin A} }
\right\rangle } \hfill \\
 {\left\langle {\varphi _{\notin A} } \mathrel{\left| {\vphantom {{\varphi _{\notin A} } {\varphi _A }}}
\right. \kern-\nulldelimiterspace} {\varphi _A }
\right\rangle } \hfill & {\left\langle {\varphi _{\notin A} } \mathrel{\left|
{\vphantom {{\varphi _{\notin A} } {\varphi _{\notin A} }}} \right.
\kern-\nulldelimiterspace} {\varphi _{\notin A} } \right\rangle } \hfill \\
\end{array} }} \right)^{-1}
\end{equation}
Here in Eq.~\ref{eqn:eq2i},
$\left\langle {\varphi } \mathrel{\left| {\vphantom {\varphi \varphi }} \right.
\kern-\nulldelimiterspace} {\varphi } \right\rangle $ is the overlap matrix
of the molecular orbitals, not an integral value. By
using the method of blockwise matrix inversion,\cite{bernstein2009matrix}
\begin{equation}
\label{eqn:eq2j}
\left( {{\begin{array}{*{20}c}
 A \hfill & B \hfill \\
 C \hfill & D \hfill \\
\end{array} }} \right)^{-1}=\left( {{\begin{array}{*{20}c}
 {\left( {A-BD^{-1}C} \right)^{-1}} \hfill & {-\left( {A-BD^{-1}C}
\right)^{-1}BD^{-1}} \hfill \\
 {-D^{-1}C\left( {A-BD^{-1}C} \right)^{-1}} \hfill & {D^{-1}C\left(
{A-BD^{-1}C} \right)^{-1}BD^{-1}+D^{-1}} \hfill \\
\end{array} }} \right)
\end{equation}
we set
\begin{equation}
\label{eqn:eq2k}
\left( {\left\langle {\varphi _A } \mathrel{\left| {\vphantom {{\varphi _A } {\varphi _A }}} \right.
\kern-\nulldelimiterspace} {\varphi _A } \right\rangle
-\left\langle {\varphi _A } \mathrel{\left| {\vphantom {{\varphi _A } {\varphi _{\notin A} }}} \right.
\kern-\nulldelimiterspace} {\varphi _{\notin A} }
\right\rangle \left\langle {\varphi _{\notin A} } \mathrel{\left| {\vphantom
{{\varphi _{\notin A} } {\varphi _{\notin A} }}} \right.
\kern-\nulldelimiterspace} {\varphi _{\notin A} } \right\rangle
^{-1}\left\langle {\varphi _{\notin A} } \mathrel{\left| {\vphantom {{\varphi _{\notin A} }
{\varphi _A }}} \right. \kern-\nulldelimiterspace} {\varphi _A }
\right\rangle } \right)^{-1}=\left\langle {\varphi _A } \right|\left( {1-\rho
_{\notin A} } \right)\left| {\varphi _A } \right\rangle ^{-1}=\alpha
\end{equation}
After some transformations, the following expressions are obtained
\begin{equation}
\label{eqn:eq2l}
 \left| {\tilde {\varphi }} \right\rangle
 =\left| {\left( {1-\rho _{\notin A} } \right)\left| {\varphi _A }
\right\rangle \alpha ,\left( {1-\left( {1-\rho _{\notin A} } \right)\left|
{\varphi _A } \right\rangle \alpha \left\langle {\varphi _A } \right|}
\right)\left| {\varphi _{\notin A} } \right\rangle \left\langle {\varphi _{\notin
A} } \mathrel{\left| {\vphantom {{\varphi _{\notin A} } {\varphi _{\notin A} }}}
\right. \kern-\nulldelimiterspace} {\varphi _{\notin A} } \right\rangle ^{-1}}
\right\rangle
\end{equation}
\begin{equation}
\label{eqn:eq2m}
 \rho =\left| {\tilde {\varphi }} \right\rangle \left\langle {\varphi _A \varphi _{\notin A} } \right|
 =\left( {1-\rho _{\notin A} } \right)\left| {\varphi _A } \right\rangle \alpha
\left\langle {\varphi _A } \right|\left( {1-\rho _{\notin A} } \right)+\rho
_{\notin A}
\end{equation}
\begin{equation}
\label{eqn:eq2n}
\frac{\partial E}{\partial \left\langle {{\varphi _A}} \right|}=4\left(
{1-\rho } \right)f\left| {\tilde {\varphi }_A } \right\rangle =4\left(
{1-\left( {1-\rho _{\notin A} } \right)\left| {\varphi _A } \right\rangle
\alpha \left\langle {\varphi _A } \right|} \right)\left( {1-\rho _{\notin A} }
\right)f\left( {1-\rho _{\notin A} } \right)\left| {\varphi _A }
\right\rangle \alpha
\end{equation}
We can further set $\alpha $ as a unit matrix to make
$\left( {1-\rho _{\notin A} } \right)\left| {\varphi _A } \right\rangle $ orthonormalized,
\begin{equation}
\label{eqn:eq2o}
\alpha =\left\langle {\varphi _A } \right|\left( {1-\rho _{\notin A} }
\right)\left( {1-\rho _{\notin A} } \right)\left| {\varphi _A } \right\rangle
^{-1}=\left\langle {\Phi _A } \mathrel{\left| {\vphantom {{\Phi _A } {\Phi
_A }}} \right. \kern-\nulldelimiterspace} {\Phi _A } \right\rangle ^{-1}=I
\end{equation}
and obtain the energy gradient for block $A$,
\begin{equation}
\label{eqn:eq2p}
\frac{\partial E}{\partial \left\langle {{\varphi _A}} \right|}=4\left(
{1-\left( {1-\rho _{\notin A} } \right)\left| {\varphi _A } \right\rangle
\left\langle {\varphi _A } \right|} \right)\left( {1-\rho _{\notin A} }
\right)f\left( {1-\rho _{\notin A} } \right)\left| {\varphi _A }
\right\rangle
\end{equation}
At the point of the lowest energy, the energy gradient is zero. We obtain the SCF equation
Eq.~\ref{eqn:blw-om} with
\begin{equation}
\label{eqn:eq2q}
\left\langle {\varphi _A } \right|\left( {1-\rho _{\notin A} } \right)f\left(
{1-\rho _{\notin A} } \right)\left| {\varphi _A } \right\rangle =\varepsilon
_A
\end{equation}
Because $\alpha $ is a unit matrix, Eq.~\ref{eqn:eq2m} now becomes
\begin{equation}
\label{eqn:eq2r}
\rho =\left( {1-\rho _{\notin A} } \right)\left| {\varphi _A } \right\rangle
\left\langle {\varphi _A } \right|\left( {1-\rho _{\notin A} } \right)+\rho
_{\notin A} =\rho _A^x +\rho _{\notin A}
\end{equation}
Eq.~\ref{eqn:eq1h} can then be obtained by putting Eq.~\ref{eqn:eq2r} into
Eq.~\ref{eqn:blw-om}.
%{\color{red}\bf?? below are some comments on the equations, moved from main text}
%but it looks complicated and
%not direct. Through our direct derivation of the energy gradient in
%the Appendix,

%Therefore, ({\color{red} \bf ??, why??, or use: Following Eq. ??, })
We have the property:
\begin{equation}
\label{eqn:eq2s}
(1 - {\rho _{ \notin A}})\left| {{\varphi _{ \notin A}}} \right\rangle
= \left| {{\varphi _{ \notin A}}} \right\rangle
- {\rho _{ \notin A}}\left| {{\varphi _{ \notin A}}} \right\rangle  = 0
\end{equation}
If we set $\left| {\Phi _i^A } \right\rangle =(1-\rho _{\notin A} )\left|
{\varphi _i^A } \right\rangle $, then $\left\langle {\varphi _{\notin A}}
\right.\left| {\Phi _i^A } \right\rangle =0$. By multiplying $\left\langle {\varphi _i^A } \right|$
on the left of Eq.~\ref{eqn:blw-o}, using the idempotency relation
$(1-\rho _{\notin A} )(1-\rho _{\notin A} )=(1-\rho _{\notin A} )$, and keeping
$\Phi _A $ orthogonal, we obtain:
\begin{equation}
\label{eqn:eq2t}
\left\langle {\Phi _i^A } \right|f\left| {\Phi _i^A } \right\rangle
=\left\langle {\Phi _i^A } \mathrel{\left| {\vphantom {{\Phi _i^A } {\Phi
_i^A }}} \right. \kern-\nulldelimiterspace} {\Phi _i^A } \right\rangle
\varepsilon _i^A =\varepsilon _i^A
\end{equation}
\begin{equation}
\label{eqn:eq2u}
f\left| {\Phi _i^A } \right\rangle =\varepsilon _i^A \left| {\Phi _i^A }
\right\rangle
\end{equation}
The above derivation from Eq.~\ref{eqn:eq2s} to Eq.~\ref{eqn:eq2u} is a reversed process
%({\color{red} \bf?? what does this mean???})
of generalized Phillips-Kleiman
pseudopotential derivation in Ref.~\citenum{weeks1968use} .
This means that the projected wavefuction of block $A$
orthogonalized to all other blocks is the eigenfuction of the whole system.

To constrain the excited state particle orbitals $\left| {{\varphi _A}} \right\rangle $
orthogonal to the ground state occupied orbitals $\left| {{\varphi _A^0}} \right\rangle $,
we need $\left\langle {{\varphi _A}} \right|\left( {1 - \rho _{ \notin A}^0} \right)
\left| {\varphi _A^0} \right\rangle {\rm{ = }}0 $. The following equation is solved with
a Lagrange multiplier $\lambda$,
\begin{equation}
\label{eqn:eq2v}
\frac{{\partial L}}{\partial \left\langle {{\varphi _A}} \right|}
= \left( {1 - \left( {1 - {\rho _{ \notin A}}} \right)\left| {{\varphi _A}} \right\rangle
\left\langle {{\varphi _A}} \right|} \right)\left( {1 - {\rho _{ \notin A}}}
\right)f\left( {1 - {\rho _{ \notin A}}} \right)\left| {{\varphi _A}} \right\rangle
- \lambda \left( {1 - \rho _{ \notin A}^0} \right)\left| {\varphi _A^0} \right\rangle {\rm{ = }}0
\end{equation}
By multiplying $\left\langle {{\varphi _A^0}} \right| $ on the left, the
Lagrange multiplier $\lambda$ can be obtained:
\begin{equation}
\label{eqn:eq2w}
\lambda  = \left\langle {\varphi _A^0} \right|\left( {1 - \left( {1 - {\rho _{ \notin A}}} \right)\left| {{\varphi _A}} \right\rangle \left\langle {{\varphi _A}} \right|} \right)\left( {1 - {\rho _{ \notin A}}} \right)f\left( {1 - {\rho _{ \notin A}}} \right)\left| {{\varphi _A}} \right\rangle
\end{equation}
The equation then becomes,
\begin{equation}
\label{eqn:eq2x}
\left( {1 - \left( {1 - \rho _{ \notin A}^0} \right)\left| {\varphi _A^0} \right\rangle \left\langle {\varphi _A^0} \right|} \right)\left( {1 - \left( {1 - {\rho _{ \notin A}}} \right)\left| {{\varphi _A}} \right\rangle \left\langle {{\varphi _A}} \right|} \right)\left( {1 - {\rho _{ \notin A}}} \right)f\left( {1 - {\rho _{ \notin A}}} \right)\left| {{\varphi _A}} \right\rangle  = 0
\end{equation}
We can further rewrite the above equation into a symmetric form,
\begin{equation}
\label{eqn:eq2y}
\rho '\left( {1 - {\rho _{ \notin A}}} \right)f\left( {1 - {\rho _{ \notin A}}} \right)\rho '
\left| {\varphi _i^A} \right\rangle  = \rho '\left( {1 - {\rho _{ \notin A}}} \right)
\left| {\varphi _i^A} \right\rangle \varepsilon _i^A
\end{equation}
where $\rho ' = 1 - \left( {1 - \rho _{ \notin A}^0} \right)\left| {\varphi _A^0} \right\rangle
\left\langle {\varphi _A^0} \right| = \left( {1 - \rho _{ \notin A}^0} \right)
\left| {\varphi _{Av}^0} \right\rangle \left\langle {\varphi _{Av}^0} \right| $
by using normalization property of complete non-orthogonal basis,
$\left| {\varphi _{Av}^0} \right\rangle $ is ground state unoccupied orbitals.
To solve this eigenequation, we multiply
$\left\langle {\varphi _{Av}^0} \right| = {\left( {\chi T_{Av}^0} \right)^\dag } $
on the left and set $\left| {{\varphi _A}} \right\rangle  = \chi {T_A} = \chi T_{Av}^0T_A'$ to
project the equation to ground state unoccupied orbitals space. The following
equation is then obtained:
\begin{equation}
\label{eqn:eq2z}
\left( {{T_{Av}}{{^0}^\dag }F_A^{'}{T_{Av}}^0} \right)T_A^{'} = \left( {{T_{Av}}{{^0}^\dag }S_A^{'}{T_{Av}}^0} \right)T_A^{'}{E_A}
\end{equation}
After the diagonalization, the particle orbitals will be selected
according to the order of ground state unoccupied orbitals.

%%%%%%%%%%%%%%%%%%%%%%
\section{Derivation of a new approximate off-diagonal matrix element
expression and its relation with previous results}
%%%%%%%%%%%%%%%%%%%%%%
According to the Kohn-Sham equation, we can obtain orthonormal
Kohn-Sham (KS) wavefunction $\left| {{\Phi ^{KS}}} \right\rangle $.
Assume $\left| {{\Phi ^{KS}}} \right\rangle $ is the wavefunction of DFT
and $H_t$ is the Harmiltonian including correlation interaction. We can obtain
$E_i^{DFT} = H_{ii}^{DFT} = \left\langle {\Phi _i^{KS}} \right|{H_t}
\left| {\Phi _i^{KS}} \right\rangle $, ${M_{ij}} = \left\langle {{\Phi _i^{KS}}}
\mathrel{\left | {\vphantom {{\Phi _i^{KS}} {\Phi _j^{KS}}}}
 \right. \kern-\nulldelimiterspace} {{\Phi _j^{KS}}} \right\rangle $ and
\begin{equation}
\label{eqn:eq3a}
H_{ii}^{DFT} = k_i^2H_{ii}^{KS} = \left\langle {{k_i}\Phi _i^{KS}} \right|H
\left| {{k_i}\Phi _i^{KS}} \right\rangle  = \left\langle {{\Psi _i}} \right|H\left| {{\Psi _i}} \right\rangle
\end{equation}
where $\left| {{\Psi _i}} \right\rangle  = \left| {{k_i}\Phi _i^{KS}} \right\rangle $
and ${k_i} = \sqrt {{{H_{ii}^{DFT}} \mathord{\left/ {\vphantom {{H_{ii}^{DFT}} {H_{ii}^{KS}}}} \right.
\kern-\nulldelimiterspace} {H_{ii}^{KS}}}} $. Similar as diagonal matrix element treatment,
the off-diagnal matrix element can be denote as
\begin{equation}
\label{eqn:eq3b}
 H_{ij}^{DFT} = \left\langle {\Phi _i^{KS}} \right|{H_t}\left| {\Phi _j^{KS}} \right\rangle
\approx \left\langle {{\Psi _i}} \right|H\left| {{\Psi _j}} \right\rangle  \\
 = {k_i}{k_j}\left\langle {\Phi _i^{KS}} \right|H\left| {\Phi _j^{KS}} \right\rangle
 = \sqrt {\frac{{H_{ii}^{DFT}H_{jj}^{DFT}}}{{H_{ii}^{KS}H_{jj}^{KS}}}} H_{ij}^{KS} \\
\end{equation}
Using first order Taylor approximation, we can obtain two previous expressions
in Ref.~\citenum{cembran2009block,zhou2017hamiltonian}.
\begin{equation}
\label{eqn:eq3c}
\begin{array}{c}
 H_{ij}^{DFT} \approx \left( {1 + \frac{{H_{ii}^{DFT} - H_{ii}^{KS}}}{{2H_{ii}^{KS}}}} \right)
\left( {1 + \frac{{H_{jj}^{DFT} - H_{jj}^{KS}}}{{2H_{jj}^{KS}}}} \right)H_{ij}^{KS} \\
\approx H_{ij}^{KS} + \frac{{H_{ij}^{KS}}}{{H_{ii}^{KS} + H_{jj}^{KS}}}
\left( {H_{ii}^{DFT} + H_{jj}^{DFT} - H_{ii}^{KS} - H_{jj}^{KS}} \right) \\
\approx H_{ij}^{KS} + \frac{{S_{ij}^{KS}}}{2}\left( {H_{ii}^{DFT} + H_{jj}^{DFT}
- H_{ii}^{KS} - H_{jj}^{KS}} \right) \\
\end{array}
\end{equation}
where $\frac{{H_{ii}^{DFT} - H_{ii}^{KS}}}{{2H_{ii}^{KS}}} \approx
\frac{{H_{ii}^{DFT} - H_{ii}^{KS}}}{{H_{ii}^{KS} + H_{jj}^{KS}}},
\frac{{H_{jj}^{DFT} - H_{jj}^{KS}}}{{2H_{jj}^{KS}}} \approx
\frac{{H_{jj}^{DFT} - H_{jj}^{KS}}}{{H_{ii}^{KS} + H_{jj}^{KS}}}$ and
$H_{ij}^{KS} \approx \frac{{S_{ij}^{KS}}}{2}\left( {H_{ii}^{KS} + H_{jj}^{KS}} \right)$
according to the extended H\"{u}ckel theory (EHT).\cite{hoffmann1963extended}
%\end{appendices}
%{\color{red}\bf end of the stuff moved from main text....}
%%%%%%%%%%%%%%%%%%%%%%%%%%%%%%%%%%%%%%%%%%%%%%%%%%%%%%%%%%%%%%%%%%%%%
%% The appropriate \bibliography command should be placed here.
%% Notice that the class file automatically sets \bibliographystyle
%% and also names the section correctly.
%%%%%%%%%%%%%%%%%%%%%%%%%%%%%%%%%%%%%%%%%%%%%%%%%%%%%%%%%%%%%%%%%%%%%
\bibliography{loc-refs-simpl}
\end{document}